\documentclass[12pt]{article}
\usepackage{hhline}
\usepackage{latexsym,graphicx}
\usepackage{subfigure}
\usepackage{amssymb}
\usepackage{amsmath}
\usepackage{amscd}
\usepackage{amsthm}
\usepackage{float}
\usepackage{xcolor}
\usepackage[left=2cm,top=2.5cm,right=2.5cm,bottom=1.5cm]{geometry}   
 
\begin{document}
\begin{center}
\large{\bf{{Investigating the hyperbolic and hybrid scalar field cosmologies with varying cosmological constant in $f(R,T)$ gravity}}} \\
\vspace{10mm}
\normalsize{Nasr Ahmed$^1$'$^2$ and Tarek M. Kamel$^2$}\\
\vspace{5mm}
\small{\footnotesize $^1$ Mathematics and statistics Department, Faculty of Science, Taibah University, Saudi Arabia.} \\
\small{\footnotesize $^2$ Astronomy Department, National Research Institute of Astronomy and Geophysics, Helwan, Cairo, Egypt\footnote{abualansar@gmail.com}} \\
\end{center}  
\date{}
\begin{abstract}
This paper investigated two scalar field cosmological models in $f(R,T)$ gravity with cosmic transit and varying cosmological constant $\Lambda(t)$.The cosmological constant tends to have a tiny positive value in the current epoch.The scalar field pressure $p_{\phi}$ shows a sign reversal for a normal scalar field. For the phantom field, the scalar potential $V(\phi)$ is negative and the energy density $\rho_{\phi}=E_k+V$ takes negative values when the equation of state parameter $\omega_{\phi}$ is less than $-1$. While the weak energy condition WEC implies that the total energy density $\rho=\sum_i\rho_i\geq 0$, we still can have a negative $\rho$ term as long as the total $\rho$ is positive. In the current work we argue that the WEC, $\rho=\sum_i \rho_i \geq 0$ and $p_i+\rho_i \geq 0$, is not violated but with an instability for the second model at late-times. For a scalar field $\phi$, The condition $\rho_{\phi}+p_{\phi}=\rho_{\phi} (1+\omega_{\phi})=2E_k\geq 0$ allows for $\rho_{\phi}<0$ if $\omega_{\phi}<-1$. The causality and energy conditions have been discussed for both models.
 The cosmology in both models was studied using a given function $a(t)$ derived from the desired cosmic behavior, which is the opposite of the traditional view.
\end{abstract}

\section{Introduction}
Accelerated cosmic expansion \cite{13,14} has become a basic motivation for a variety of modified gravitational theories \cite{noj1}-\cite{nass222}. In order to find a satisfactory explanation, an exotic form of energy with negative pressure, called dark energy, was hypothesized. Several dynamical scalar fields models of dark energy introduced such as the Quintessence, Phantom and Tachyons \cite{quint}-\cite{ark22}. For a zero curvature FRW universe driven by a scalar field $\phi$, Einstein’s equations are
\begin{equation}
3H^2=\frac{1}{2} \dot{\phi}^2+V(\phi) ~~~,~~~\dot{H}=-\frac{1}{2} \dot{\phi}^2~~~,~~~\ddot{\phi}+3H\dot{\phi}+V'=0 , 
\end{equation}
With units $8\pi M_{Pl}^{-2}=c=1$. $H=\frac{\dot{a}}{a}$ is the Hubble parameter and $V(\phi)$ is the potential. The prime denotes differentiation with respect to $\phi$, and the dots denote differentiation with respect to $t$. While this nonlinear system is insoluble in general, a progress can be made through postulating a particular form of the scale factor $a(t)$ and then get the form of both $\phi(t)$ and $V(\phi)$ \cite{barrow,ellis}. 
In \cite{barenjee}, it has been shown that a minimally coupled scalar field in Brans-Dicke theory leads to an accelerating universe. A power function forms of the scale factor $a$ and the scalar field $\phi$ were assumed as 
\begin{equation}
a(t)=a_1t^{\alpha} ~~~~~,~~~~~\phi(t)=\phi_1 t^{\beta},
\end{equation}
with $a_1$, $\phi_1$, $\alpha$ and $\beta$ are constants. An accelerated expansion was also  achieved in the modified Brans-Dicke theory by considering the follwoing power-law form of both $a$ and $\phi$ \cite{martin}.
\begin{equation}
a(t)= a_0\left(\frac{t}{t_0}\right)^{\alpha},~~~~~\phi(t)=\phi_0\left(\frac{t}{t_0}\right)^{\beta}.
\end{equation}
Cosmology in the scalar-tensor $f(R,T)$ gravity has been studied in \cite{rosa} where three particular forms of $a(t)$ have been used.
\subsection{Negative potentials and energy densities}
The case of negative potential cosmologies has become interesting after the prediction of Ads spaces in string theory and particle physics. Negative potentials also exist in ekpyrotic and cyclic cosmological models in which the universe goes from a contracting to an accelerating phase \cite{cyc,ekp}. They are commonly predicted in particle physics, supergravity and string theory where the general vacuum of supergravity has a negative potential. It has also been suggested that negative potentials lead to an explanation of the cosmological scale in terms of a high energy scale such as the supersymmetry breaking scale or the electroweak scale \cite{scale}. A detailed discussion of scalar field cosmology with negative potentials was carried out in \cite{linde}. The impact of negative energy densities on classical FRW cosmology has been investigated in \cite{-ve} where the total energy density can be expanded as 
\begin{equation}
\rho=\sum_{n=-\infty}^{\infty}\rho_n^+a^{-n}+\sum_{m=-\infty}^{\infty}\rho_m^-a^{-m},
\end{equation}
where $\rho_n^+$ is the familiar positive energy density and $\rho_m^-$ is the negative cosmological energy density. Cosmic evolution with negative energy densities was also examined in \cite{-ve2} where vacuum polarization was mentioned as an example  of a gravitational source with $\rho<0$ that may have played a significant role in early cosmic expansion.\par
An interesting study was carried out in \cite{mexico} where the equation of state parameter is negative ( $\omega_{\phi}=p_{\phi}/\rho_{\phi}<-1$ ) with no violation of the weak energy condition ($\rho=\sum_i \rho_i \geq 0$ \& $p_i+\rho_i \geq 0$) which requires a negative potential $V(\phi)<0$. It has been shown that $\rho_{\phi}=\frac{1}{2}\dot{\phi}^2+V(\phi)$ becomes negative with $\omega_{\phi}<-1$, the negative $\rho_{\phi}$ leads to a small value of the cosmological constant. However, while cosmic expansion exists in such scenario, the negative potential $V$ leads to a collapsing universe.\par
The classical energy conditions are ``the null energy condition (NEC) $\rho + p\geq 0$; weak energy condition (WEC) $\rho \geq 0$, $\rho + p\geq 0$; strong energy condition (SEC) $\rho + 3p\geq 0$ and dominant energy condition (DEC) $\rho \geq \left|p\right|$''. Since the SEC implies that gravity should always be attractive, this condition fails in the acceleration and inflation epochs \cite{ec3,ec4}. As mentioned in \cite{clec}, even the simplest scalar field theory we can write down violates the SEC. The NEC is the most fundamental energy condition on which the singularity theorems, and other key results, are based \cite{necvb}. If the NEC is violated, all other point-wise energy conditions (ECs) are automatically violated. A very useful discussion about the validity of classical linear ECs was given in \cite{clec} where it has been shown that these classical conditions can not be valid in general situations. 
The scalar field potential $V(\phi)$ is restricted by the ECs where the scalar field $\phi$ ( with $\rho_{\phi}=\frac{1}{2} \dot{\phi}^2+V(\phi)$ \& $p_{\phi}=\frac{1}{2} \dot{\phi}^2-V(\phi)$ ) satisfies the NEC for any $V(\phi)$, the WEC if and only if $V(\phi)\geq -\frac{1}{2} \dot{\phi}^2$, the DEC if and only if $V(\phi)\geq 0$, the SEC if and only if $V(\phi)\leq \dot{\phi}^2$. The detailed proof of this theorem can be found in \cite{thesis}.
\subsection{$\Lambda(t)$ models}
A new model for the time-dependent cosmological constant $\Lambda(t)$ was proposed in \cite{21} using the following ansatz
\begin{equation} \label{cosma1}
\Lambda =  \frac{\Lambda_{Pl}}{\left(t/t_{Pl}\right)^2} \propto \frac{1}{t^2},
\end{equation}
$\Lambda$ starts at the Planck time as $\Lambda_{Pl} = \sim M_{Pl}^2$ and leads to the value $\Lambda_{0} \sim 10^{-120} M_{Pl}^2$ for the current epoch. The decay of $\Lambda(t)$ during inflation and as Bose condensate evaporation was studied in \cite{kholopov1,kholopov2}. Other models for $\Lambda(t)$ have been suggested in \cite{20d,20b,20c,20cc,20cc1}. The following ansatz was first introduced in \cite{20d} where a variety of cosmologically relevant
observations were used to put impose strict constraints on $\Lambda(t)$ models
\begin{equation} \label{vary2}
\Lambda(H)= \lambda +\alpha H + 3 \beta H^2 ,
\end{equation}
where $H$ is the Hubble parameter, $\lambda$, $\alpha$ and $\beta$ are constants. It has been found in \cite{20b, 20f, 20g, 20h} that the zero value of $\lambda$ doesn't agree with observations, while $\lambda \neq 0$ behaves like $\Lambda$CDM model at late-time. Examples of varying $\Lambda$ models in terms of the Hubble parameter $H$ are \cite{20b}
\begin{eqnarray} 
\Lambda(H)= \beta H +3H^2 + \delta H^n,\,\,\,\,\,\, n \in R-\left\{0,1\right\} ,\\
\Lambda(H, \dot{H}, \ddot{H})=\alpha+\beta H+\delta H^2+ \mu \dot{H}+\nu \ddot{H}.
\end{eqnarray}
A generalized holographic model of dark energy in which the effective cosmological constant depends on $H$ and its derivatives was proposed in \cite{Nojiri:2021iko,Nojiri:2020wmh,Nojiri:2021jxf}.
\subsection{$f(R,T)$ modified gravity}
The action of $f(R,T)$ modified gravity is given as \cite{frt}
\begin{equation}
\label{eq1}S=\int \left(\frac{f(R,T)}{16\pi G}+L_m\right)\sqrt{-g}~d^{4}x , 
\end{equation}
where $L_{m}$ is the matter Lagrangian density. $f(R,T)$ is an arbitrary function of the Ricci scalar $R$ and the trace $T$ of the energy-momentum tensor $T_{\mu \nu}$ defined as
\begin{equation}
\label{eq3}T_{\mu \nu}=g_{\mu \nu} L_{m}-2\frac{\partial L_{m}}{\partial g^{\mu \nu}}.
\end{equation}
Varying the action (\ref{eq1}) gives 
\begin{eqnarray}
\label{eq4} f_{R}(R,T)R_{\mu \nu}-\frac{1}{2} f(R,T)g_{\mu \nu}+(g_{\mu \nu} \Box  -\nabla_{\mu} \nabla_{\nu})f_{R}(R,T) \\    \nonumber
=8\pi T_{\mu \nu}-f_{T}(R,T)T_{\mu \nu}-f_{T}(R,T)\Theta_{\mu\nu},
\end{eqnarray}
where $\Box = \nabla^{i}\nabla_{i}$, $f_{R}(R,T)=\frac{\partial f(R,T)}{\partial R}$, $f_{T}(R,T)=\frac{\partial f(R,T)}{\partial T}$ 
and $\nabla_i$ denotes the covariant derivative. $\Theta_{\mu \nu}$ is given by
\begin{equation}
\label{eq5}\Theta_{\mu \nu}=-2T_{\mu\nu}+g_{\mu\nu}L_{m}-2g^{\alpha \beta}\frac{\partial^{2}L_{m}}{\partial g^{\mu\nu}\partial g^{\alpha \beta}} .
\end{equation}
The cosmological equations for $f(R,T)=R+2h(T)$ with cosmological constant $\Lambda$ considering a scalar field $\phi$ coupled to gravity have been given in \cite{sahoo} as 
\begin{eqnarray} \label{eq111}
\frac{2\ddot{a}}{a}+\frac{\dot{a}^2}{a^2}=4\pi \epsilon \dot{\phi}^2-8\pi V(\phi) +\mu \epsilon \dot{\phi}^2- 4\mu V(\phi)-\Lambda , \\
\frac{3\dot{a}^2}{a^2}= -4\pi \epsilon \dot{\phi}^2-8\pi V(\phi) -\mu \epsilon \dot{\phi}^2- 4\mu V(\phi)-\Lambda,\label{eq222}
\end{eqnarray}
where $h(T)=\mu T$ and $\mu$ is a constant. $\epsilon =\pm 1$ corresponding to normal and phantom scalar fields respectively. 
In the current work, two cosmological models in modified $f(R,T)$ gravity were examined using a given scale factor $a(t)$ derived from the desired cosmic behavior which is the opposite of the conventional viewpoint. Such ad hoc approach to the cosmic scale factor and cosmological scalar fields has been widely used by many authors in various theories \cite{ellis,el2,sen,senta, sent,sent11,sz,sch,ric,nasrtarek,extra,Nojiri:2022xdo}. We will make use of the following hyperbolic and hybrid scale factors:
\begin{equation} \label{sols}
a(t)= A \sinh^{\frac{1}{n}}(\eta t) ~~~~~,~~~~~a(t)=a_1 t^{\alpha_1} e^{\beta_1 t},
\end{equation}
 Where $A$, $\eta$, $n$, $a_1>0$, $\alpha_1 \geq 0$ and $\beta_1 \geq 0$ are constants. The first scale factor generates a class of accelerating models for $n > 1$, the models also exhibit a phase transition from the early decelerating epoch to the present accelerating era in a good agreement with recent observations.  The second hybrid ansatz is a mixture of power-law and exponential-law cosmologies, and can be regarded as a generalization to each of them. The power-law cosmology can be obtained for $\beta_1=0$, and the exponential-law cosmology can be obtained for $\alpha_1=0$. New cosmologies can be explored for $\alpha_1>0$ and $\beta_1>0$. A generalized form of the hybrid scale factor has been proposed in \cite{Nojiri:2022xdo,Odintsov:2021yva} to unify the cosmic evolution of the universe from a non-singular bounce to the viable dark energy
\begin{equation} 
a(t)= \left[1+a_0\left(\frac{t}{t_0}\right)^2\right]^{\frac{1}{3(1+\omega)}} \exp\left[\frac{1}{(\alpha-1)}\left(\frac{t_s-t}{t_0}\right)^{1-\alpha}\right],
\end{equation}
where $\omega$, $\alpha$ and $t_s$ are various parameters. Setting $t_0=1$ billion years, this can be re-written as the product of two scale factors
\begin{equation} 
a(t)= \left[1+a_0 t^2\right]^{\frac{1}{3(1+\omega)}} \times \exp \left[\frac{1}{(\alpha-1)}\left(t_s-t\right)^{1-\alpha}\right].
\end{equation}
In the current work, we are going to use the ansatz (\ref{vary2}) for the time varying cosmological constant which which leads to a very tiny positive value of $\Lambda$ at the present epoch as suggested by observations \cite{1,1a}. 
\section{Model 1}
\begin{figure}[H] \label{tap1}
  \centering             
  \subfigure[$q$]{\label{F1}\includegraphics[width=0.29\textwidth]{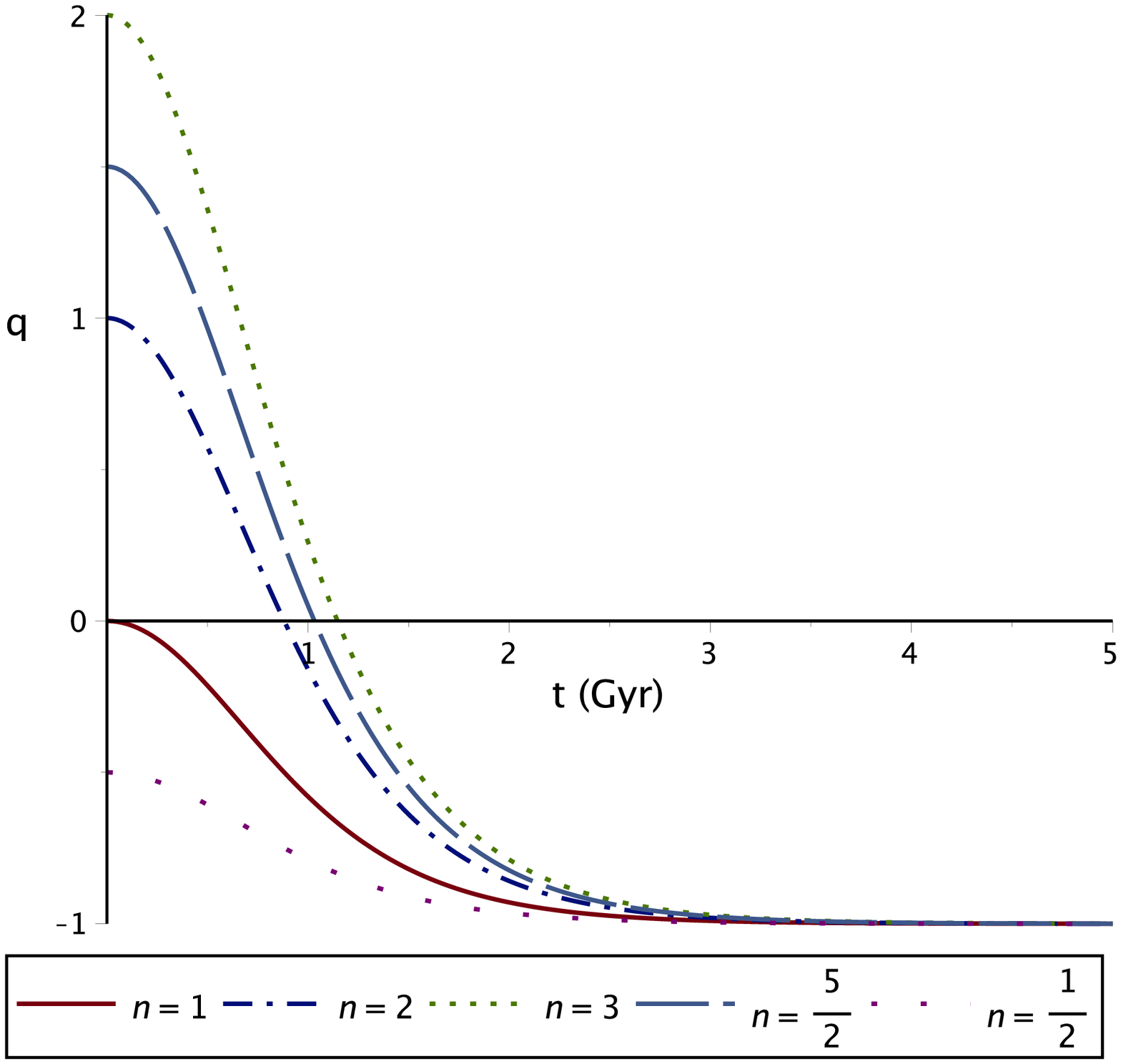}} 
	\subfigure[$j$]{\label{F2}\includegraphics[width=0.29\textwidth]{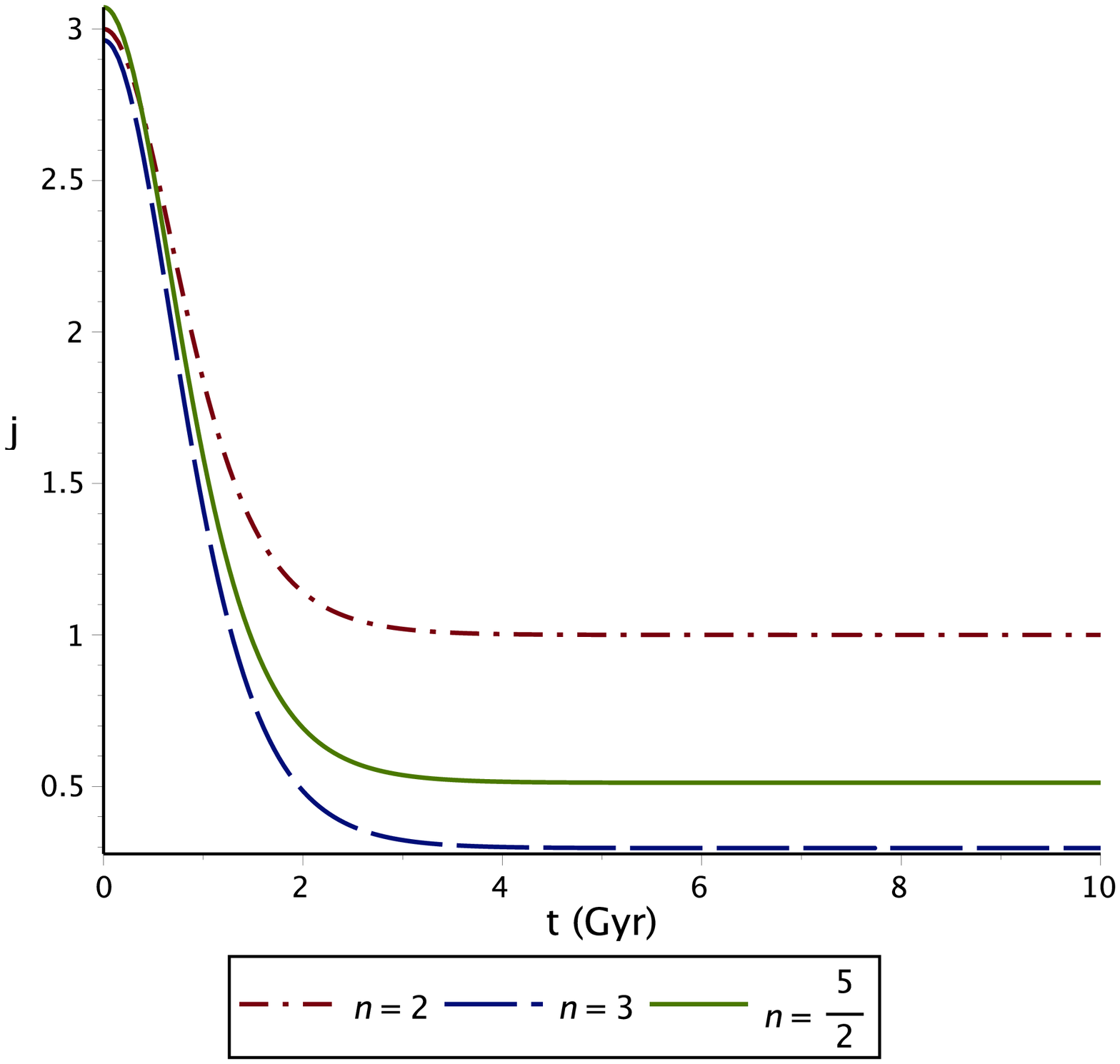}} 
	  \subfigure[$\Lambda(t)$]{\label{F3}\includegraphics[width=0.29\textwidth]{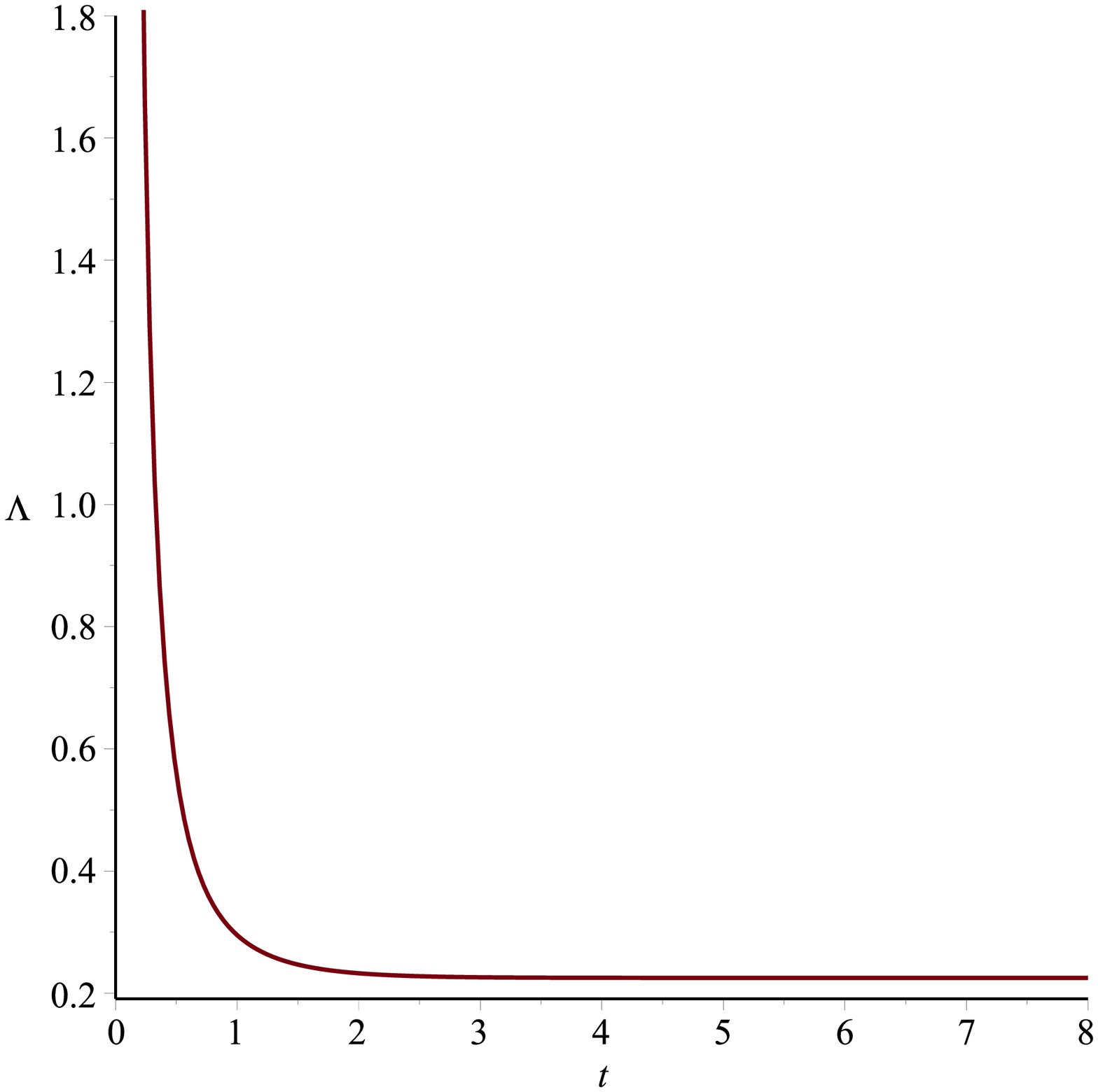}} \\
			  \subfigure[$p_{\phi}$]{\label{F4}\includegraphics[width=0.29 \textwidth]{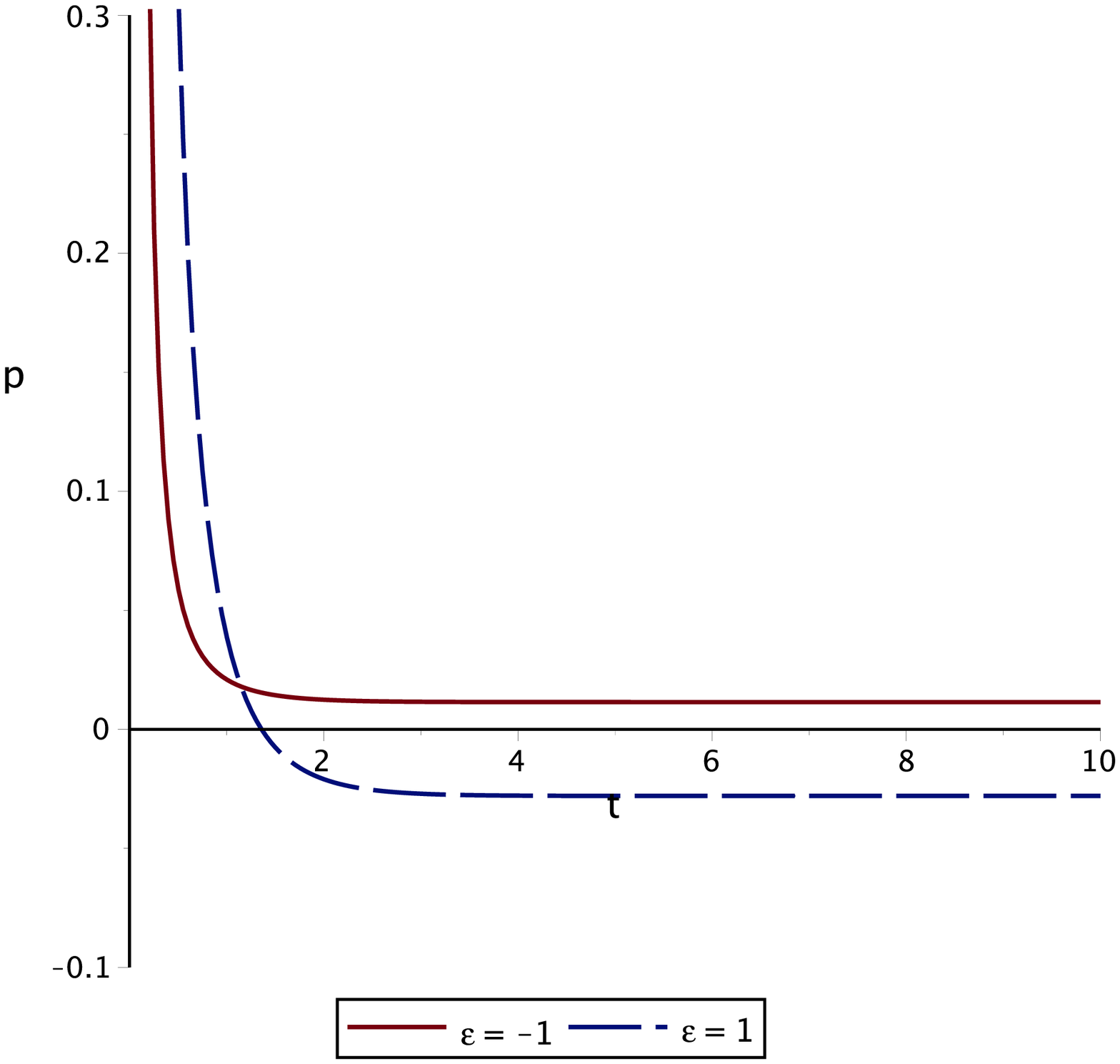}}
				\subfigure[$\rho_{\phi}$]{\label{F5}\includegraphics[width=0.29 \textwidth]{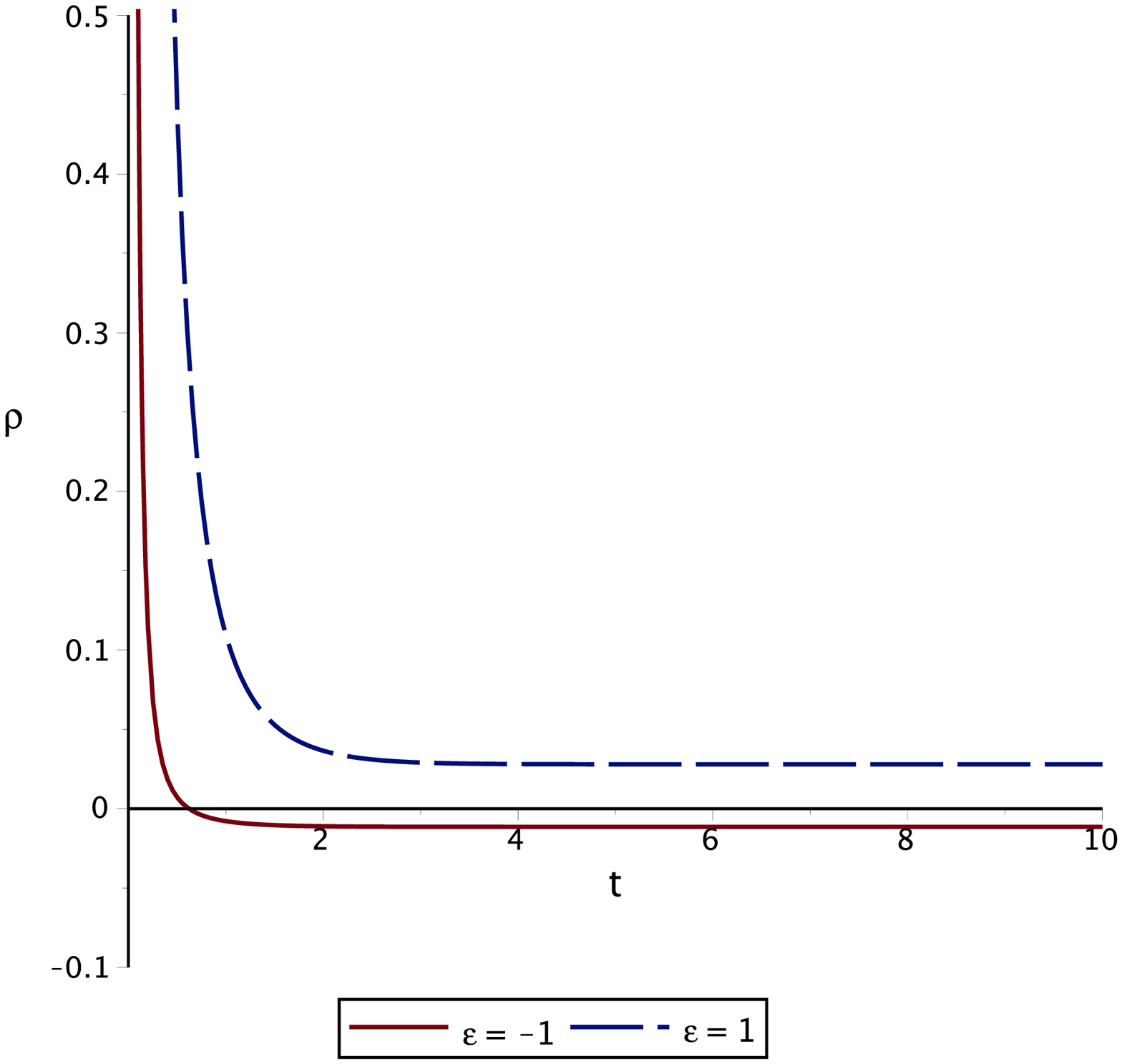}}
	\subfigure[$\omega_{\phi}(t)$]{\label{F7}\includegraphics[width=0.29\textwidth]{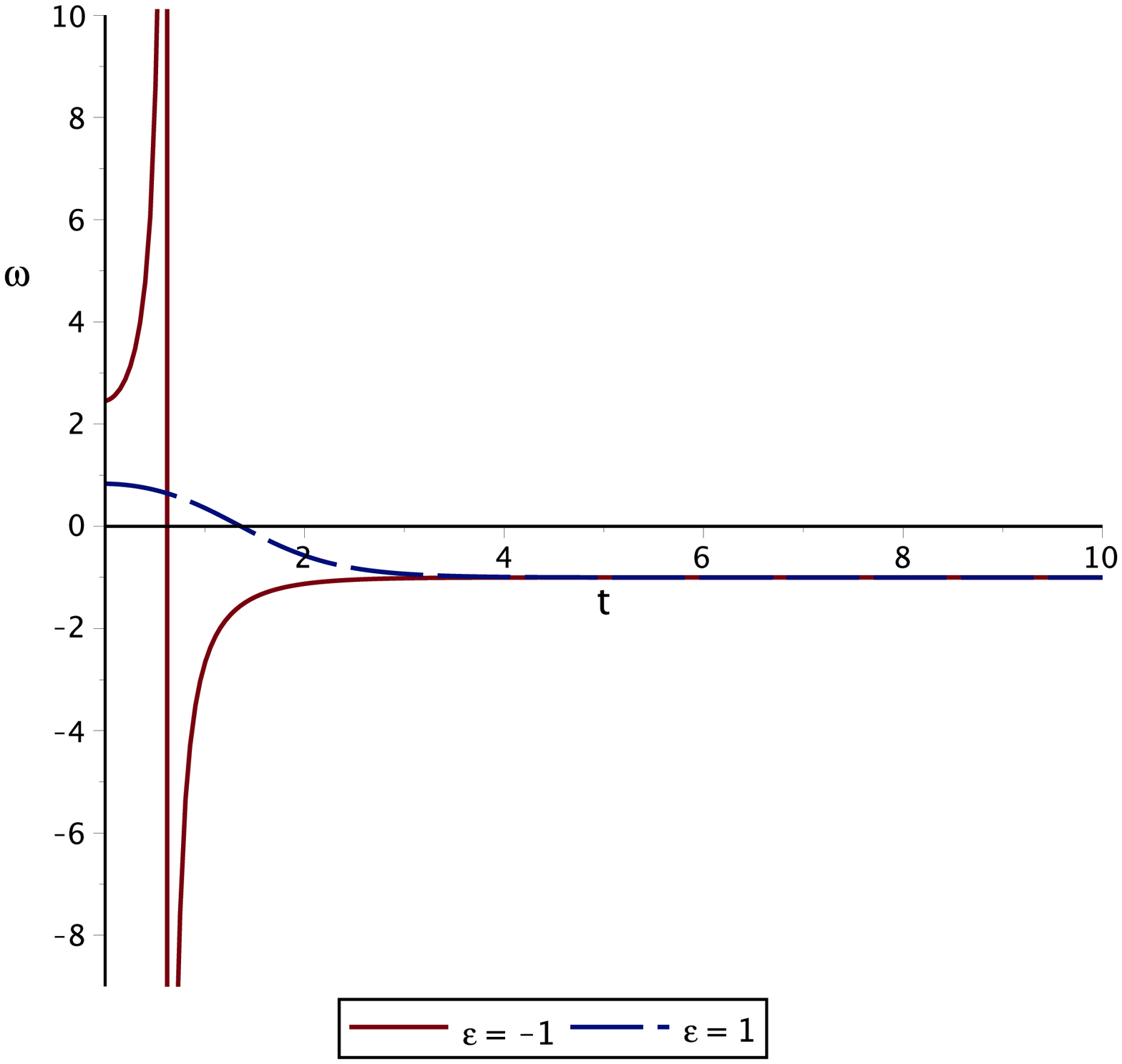}}  \\
	\subfigure[$\phi(t)$]{\label{F8}\includegraphics[width=0.29\textwidth]{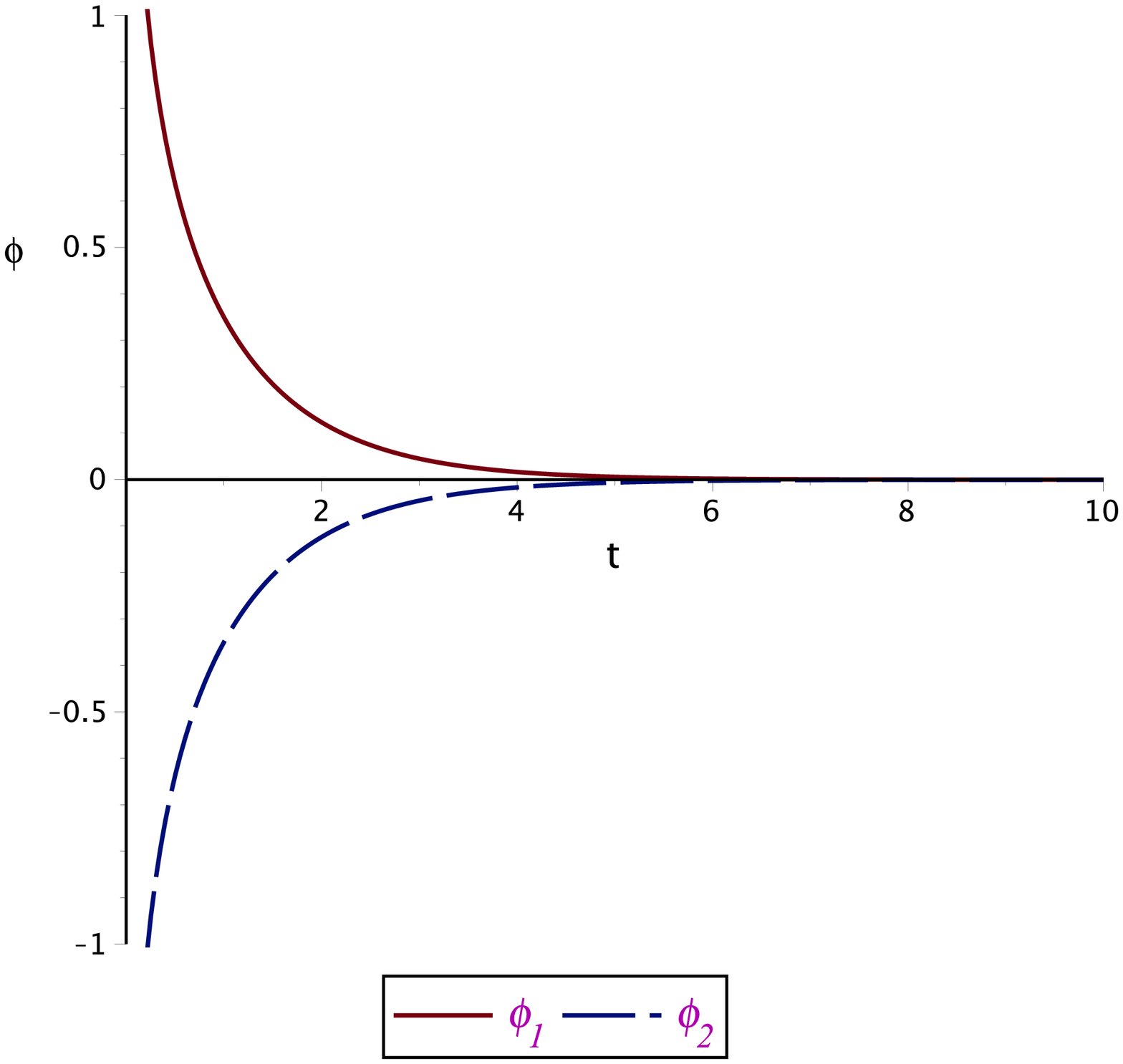}} 
		\subfigure[$V(t)$]{\label{F81}\includegraphics[width=0.29\textwidth]{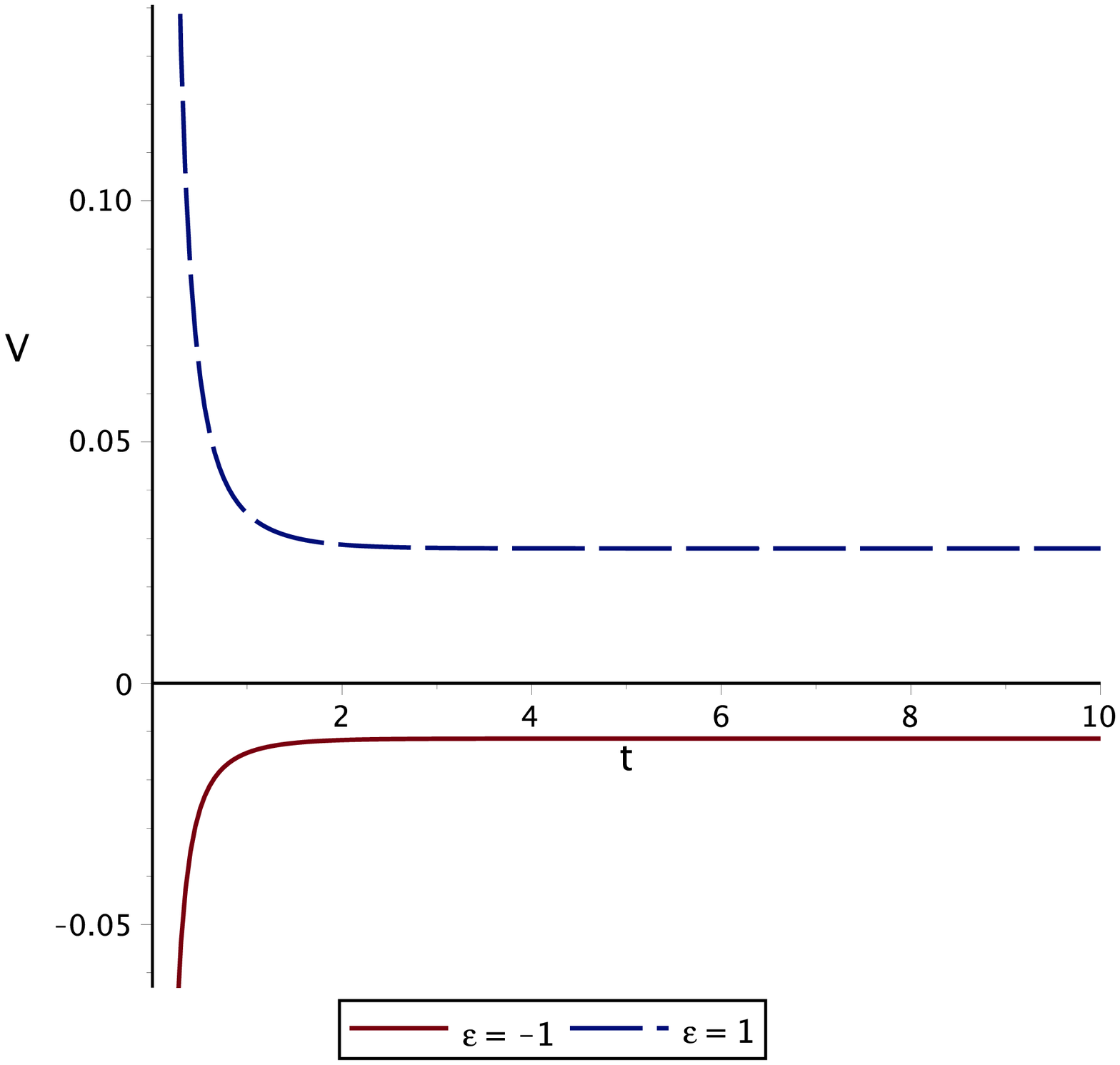}}
			\subfigure[$V(\phi)$]{\label{F9}\includegraphics[width=0.29\textwidth]{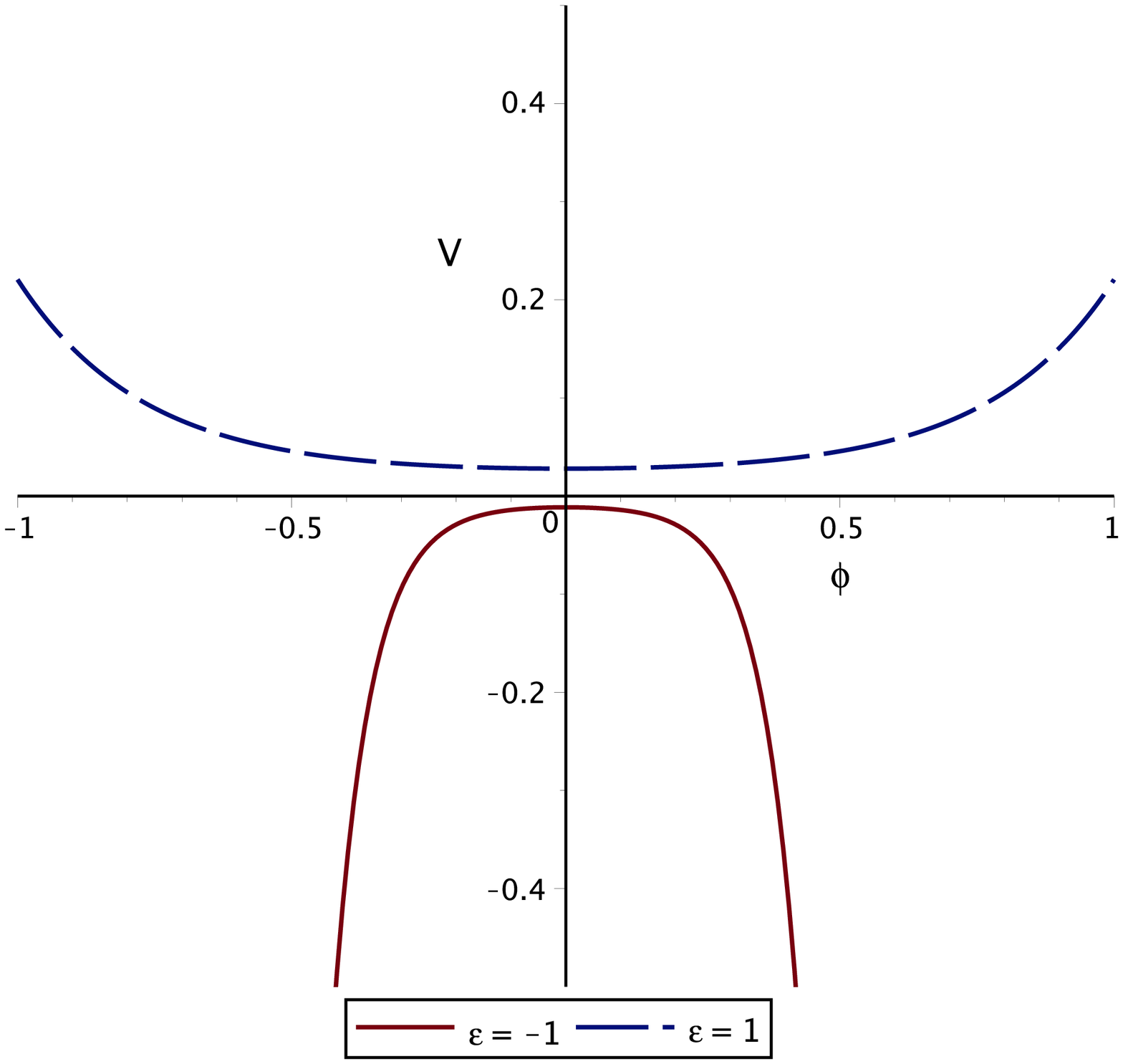}}
  \caption{The hyperbolic solution: (a) The deceleration parameter $q$ shows a decelerating-accelerating cosmic transit. (b) The jerk parameter approaches unity at late-times where the model tends to a flat $\Lambda$CDM model. (c) The cosmological constant reaches a very tiny positive value at the current epoch. (d ), (e) \& (f) show $p_{\phi}$, $\rho_{\phi}$ and $\omega_{\phi}$ for $\epsilon=\pm 1$. For the phantom case, the energy density $\rho_{\phi}=E_k+V<0$ when $\omega_{\phi}<-1$. (g) The two solutions of $\phi(t)$ obtained in (\ref{phii}). (h) The scalar potential evolution with time. (g) scalar potential $V$ verses $\phi$ . Here $ n=2, \eta=1, \phi_0=0, A=\lambda=\beta=\alpha=0.1$, $\mu=15$ for $\epsilon=-1$ and $-15$ for $\epsilon=1$.}
  \label{fig:1}
\end{figure}

Starting with the hyperbolic solution in (\ref{sols}), which gives the desired behavior of the deceleration and jerk parameters, we obtain the Hubble, deceleration, and jerk parameters as:
\begin{equation} \label{q1}
H=\frac{\eta}{n}\coth(\eta t),~~~q=-\frac{\ddot{a}a}{\dot{a}^2}=\frac{-\cosh^2(\eta t)+n}{\cosh^2(\eta t)},~~~ j=\frac{\dddot{a}}{aH^3}= 1+\frac{2n^2-3n}{\cosh^2(\eta t)}.
\end{equation}
In order to solve the system of equations (\ref{eq111}) and (\ref{eq222}) for the scalar field and the potential, we utilize the hyperbolic scale factor in (\ref{sols}) along with the time-dependent anstaz for the cosmological constant (\ref{vary2}). Then, we will have a system of two equations in two unknowns which we have solved using Maple software and obtained
\begin{eqnarray} \label{phipot}
\phi(t)&=&\frac{\mp\ln(e^{\eta t} +1)\pm\ln(e^{\eta t} -1)}{\sqrt{-2\epsilon(4\pi+\mu)}}+\phi_0  , \label{phii}\\
V(t)&=&-\frac{\left(\eta^2(1+3\beta)\coth^2(\eta t)+2\eta\alpha\coth(\eta t)+ 2(\eta^2+4\lambda)\right)}{16(2\pi+\mu)} ,\\
V(\phi)&=&-\frac{\left( (3\beta+1) \eta^2 \chi^2+4\eta \alpha\chi +2\eta^2(3\beta+5)+16\lambda+
4\eta\alpha \chi^{-1}+\eta^2\chi^{-2}(3\beta+1)\right)}{64(2\pi+\mu)} ,\label{Vphi}
\end{eqnarray}
Where $\chi \equiv e^{(\phi_0-\phi)\sqrt{-2\epsilon(4\pi+\mu)}}$ and we have used $t(\phi)= \frac{1}{\eta}\ln(\mp\frac{1+\chi}{\chi-1})$ to get the expression for $V(\phi)$. The expression for $\phi(t)$ shows that $\epsilon$ can be $-1$ provided that $(4\pi+\mu)>0$, and it can be $+1$ provided that $(4\pi+\mu)<0$. Plotting $t(\phi)$ leads to same graph for both signs in (\ref{phii}). We also obtain same expressions for $V(\phi)$ (\ref{Vphi}), energy density $\rho$ and pressure $p$ for both $\phi$ solutions. Actually, Figure\ref{F8} shows that both solutions for $\phi$, although they have a different start, unite in one solution. We can use $\phi_0=0$ without loss of generality. Recalling that $\rho_{\phi}=E_k+V$ and $p_{\phi}=E_k-V$ we obtain 
\begin{equation} \label{pandrho}
p_{\phi}(t)= -\frac{\eta^2 e^{2\eta t}}{\epsilon(4\pi+\mu)(e^{\eta t}+1)^2(e^{\eta t}-1)^2}-V(t) ~,~\rho_{\phi}(t)=-\frac{\eta^2 e^{2\eta t}}{\epsilon(4\pi+\mu)(e^{\eta t}+1)^2(e^{\eta t}-1)^2}+V(t) .
\end{equation}
The evolution of the cosmological constant in this work agrees with observations where it has a very tiny positive value at the present epoch (Figure \ref{F3}). The expressions for the parameters $q$, $j$ and the cosmological constant in equation (\ref{vary2}) are all independent of $\epsilon$. The rest of parameters are all plotted for $\epsilon=\pm 1$. 
For $\epsilon=+1$, which corresponds to normal scalar field, the scalar field pressure $p_{\phi}$ changes sign from positive to negative. We can also see that $V(\phi)$, $V(t)$ and $\rho_{\phi}$ are all positive where both $V(t)$ and $\rho_{\phi}$ tend to $\infty$ as $t \rightarrow 0$. For $\epsilon=-1$, which corresponds to phantom scalar field, the pressure $p_{\phi}>0$ all the time while $\rho_{\phi}$ takes negative values when $\omega_{\phi}<-1$ with a negative scalar potential $V$. In the literature, it is known that the vacuum phantom energy has some unusual physical properties such as the increasing vacuum energy density, violation of the DEC $\rho+p<0$ and the superluminal sound speed \cite{phan}.\par
According to the WEC, the total energy density and pressure should follow the inequalities $\rho+p=\rho(1+\omega)\geq 0$ and $\rho \geq 0$. For a scalar field $\phi$, The condition $\rho_{\phi} + p_{\phi}=\rho_{\phi} (1+\omega_{\phi})=2E_k\geq 0$ allows for $\rho_{\phi}<0$ if $\omega_{\phi}<-1$ as long as the total energy density $\rho \geq 0$ with the total equation of state parameter $\omega>-1$. In general, the phantom energy doesn't obey the WEC where it has $\rho_{ph}>0$ but $\rho_{ph} + p_{ph}=\rho_{ph} (1+\omega_{ph})=2E_k < 0$ which means that the phantom field has a negative (noncanonical) kinetic term \cite{mexico}.
Testing the classical energy conditions \cite{ec4} shows that both the null and the dominant are satisfied all the time. The highly restrictive SEC $\rho + 3p\geq 0$ is violated as expected where we have a source of repulsive gravity represented by the negative pressure which can accelerate cosmic expansion. Because the strong condition implies that gravity should always be attractive, it's expected to be violated during any accelerating epoch dominated by a repulsive gravity effect such as cosmic inflation. In addition to the ECs, the sound speed causality condition $0 \leq \frac{dp}{d\rho} \leq 1$ is satisfied only for $\epsilon=+1$.\par
The possible values of the parameters in the Figures are restricted by observations where the theoretical model should predict the same behavior obtained by observations. For that reason, we have to fine-tune the parameters' values to agree with observational results.  We have taken $n=2$ as it allows for a decelerating-accelerating cosmic transit and also allows the jerk parameter $j$ to approaches unity at late-times in an agreement with the standard $\Lambda CDM$ model. The constants $A$, $\eta$, and the integration constant $\phi_0$ are arbitrary and we have chosen the values $0.1$, $1$ and $0$ respectively without loss of generality. The value of the constant $\mu$ has been adjusted such that the quantity under the quadratic root in (\ref{phipot}) is always positive for both normal and phantom fields. If we choose $\mu=15$, then $(4\pi+\mu) > 0$ for the normal field where $\epsilon = +1$. For the phantom field with $\epsilon = -1$, we choose $\mu=-15$ so  $(4\pi+\mu) < 0$ and then $-2\epsilon (4\pi+\mu) > 0$. As we have indicated in section 1.2, the zero value of $\lambda$ doesn't agree with observations while $\lambda \neq 0$ behaves like $\Lambda CDM$ model at late-time. Based on this, we have chosen the non-zero value $0.1$ for $\lambda$, $\beta$ and $\alpha$.
\begin{figure}[H] \label{tap12}
  \centering             
  \subfigure[$\epsilon=+1$]{\label{p110}\includegraphics[width=0.29\textwidth]{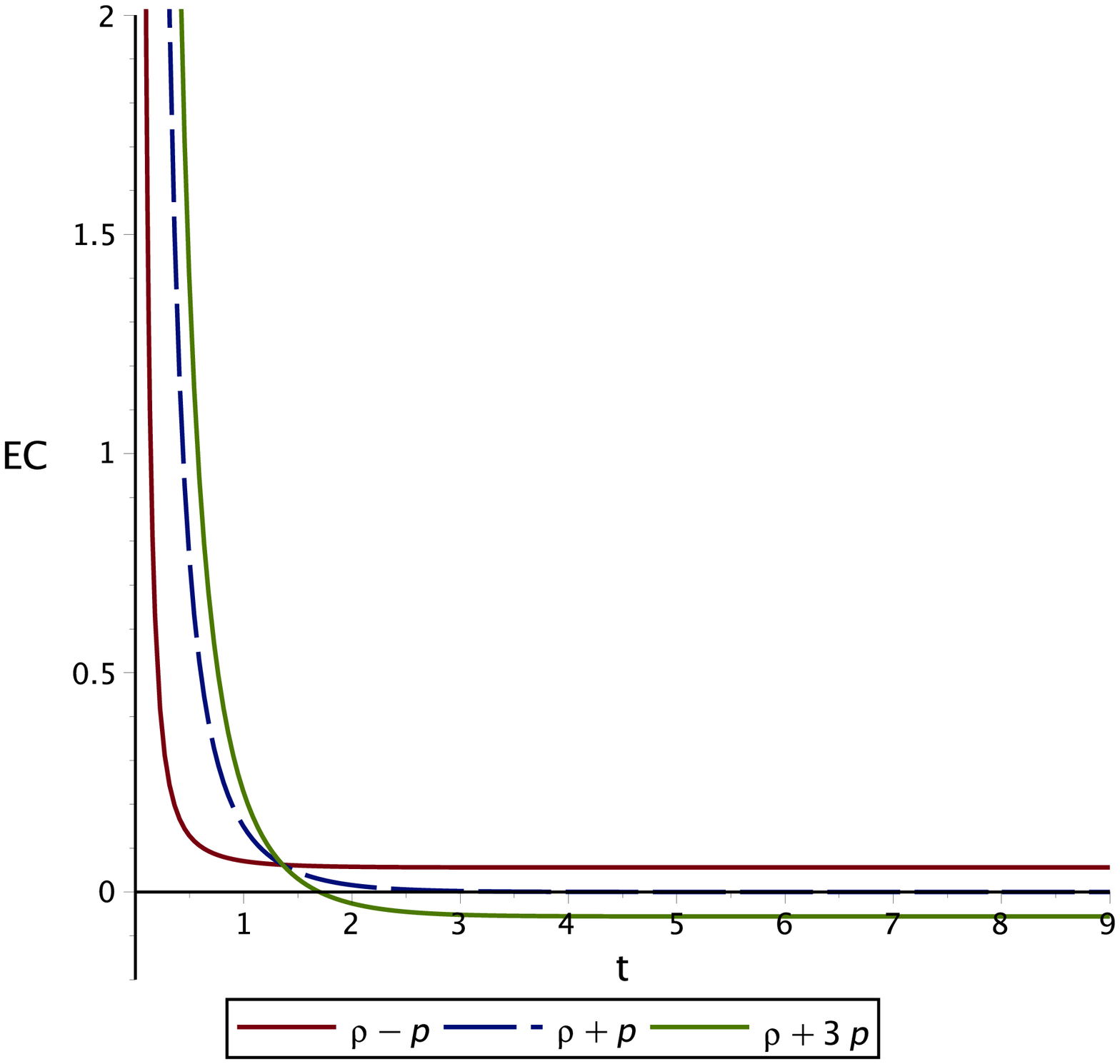}} 
		\subfigure[$\epsilon=-1$]{\label{p125}\includegraphics[width=0.29\textwidth]{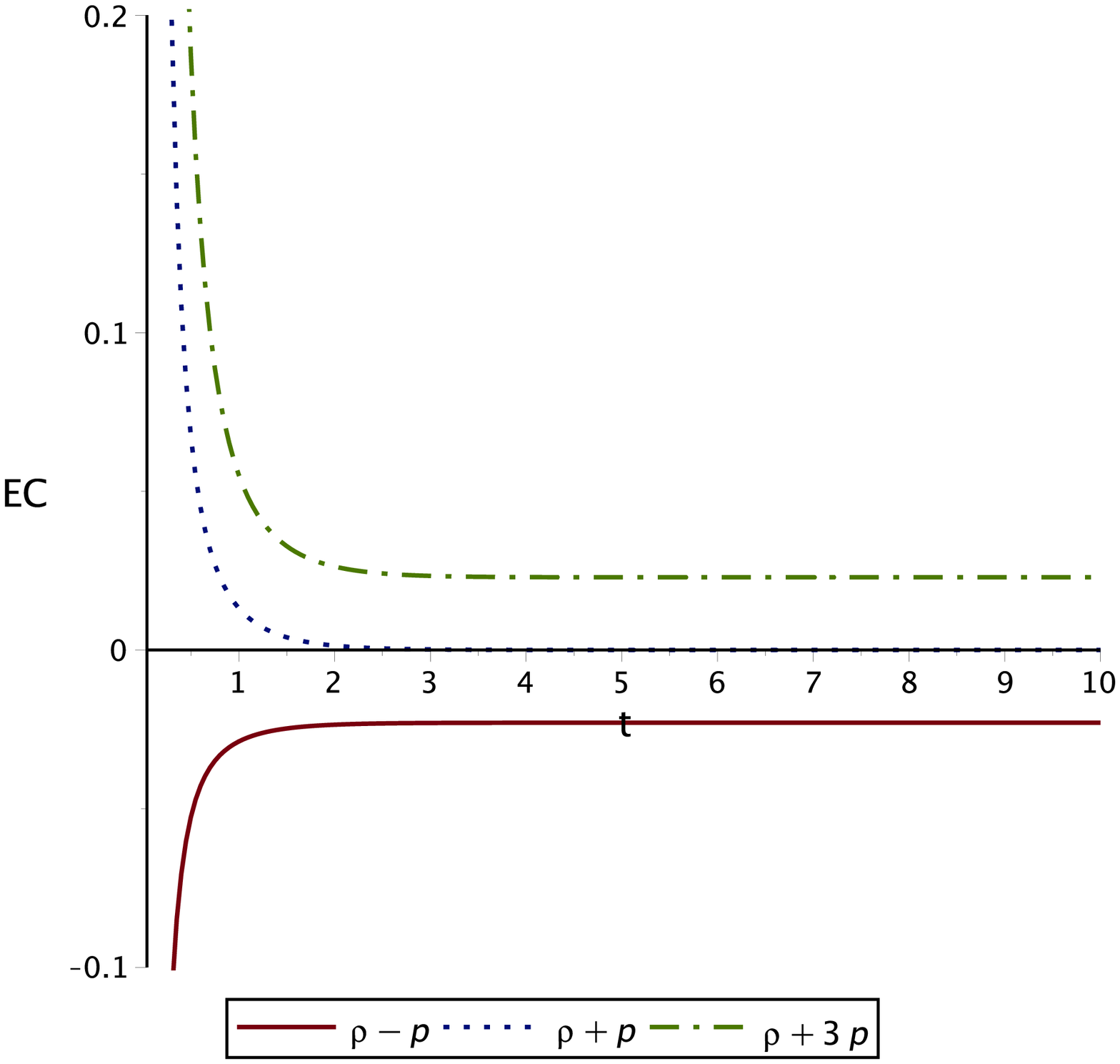}} 
	\subfigure[$dp/dt$]{\label{p120}\includegraphics[width=0.29\textwidth]{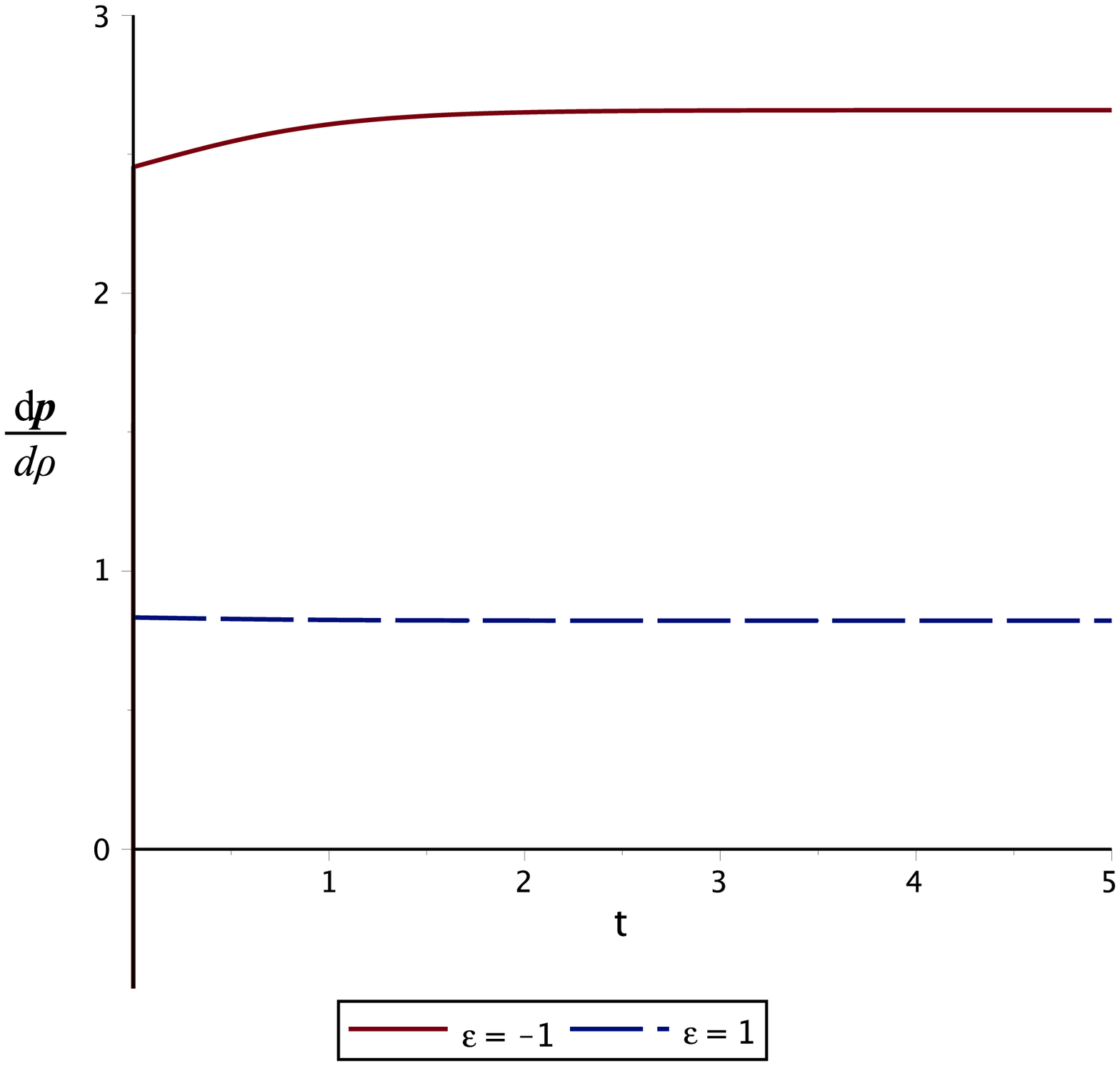}} 
  \caption{ECs and sound speed for the hyperbolic model. Superluminal sound speed for the phantom field.}
  \label{fig:31}
\end{figure}

\section{Model 2}
Considering the second hybrid scale factor in (\ref{sols}), which also leads to the desired behavior of both $q$ and $j$ \cite{ric}, we get the expressions for  $H$, $q$ and $j$ as:
\begin{equation} \label{q2}
H=\beta_1+\frac{\alpha_1}{t}~~,~~q=\frac{\alpha_1}{(\beta_1t+\alpha_1)^2}-1 ~~ ,~~  j={\frac {{\alpha_1}^{3}+ \left( 3\,\beta\,t-3 \right) {\alpha_1}^{2}+\left( 3\,{\beta}^{2}{t}^{2}-3\,\beta\,t+2 \right) \alpha_1+{\beta}^{3}
{t}^{3}}{ \left( \beta\,t+\alpha_1 \right) ^{3}}}.
\end{equation}
For the scalar field and the potential, making use of (\ref{vary2}), we get
\begin{eqnarray} \label{phipot2}
\phi(t)&=& \pm \frac{\sqrt{-\epsilon(4\pi+\mu)\alpha_1}\ln t}{\epsilon(4\pi+\mu)}+C_1,\\
V(t)&=& \frac{((3\beta_1^2(\beta_0+1)+\alpha_0\beta_1+\lambda_0)t^2+(6\alpha_1\beta_1(\beta_0+1)+\alpha_0\alpha_1)t+3\alpha_1^2(\beta_0+1)+\alpha_1}{-4(\mu+2\pi)t^2},\\
V(\phi)&=& \frac{3(\beta_1^2+\alpha_1^2)(\beta_0+1)+\alpha_0\beta_1+ \xi^{-1}\alpha_1(6\beta_0\beta_1+\alpha_0+6\beta_1)+\lambda_0-\alpha_1}{-4(2\pi+\mu)} ,
\end{eqnarray}
Where $\xi =e^{\frac{\epsilon(|C_1-\phi)(4\mu+\pi)}{\sqrt{-\epsilon\alpha_1(4\mu+\pi)}}}=t(\phi)$. Plotting $t(\phi)$ leads to same graph for both signs. Also, both solutions for $\phi$ gives the same expressions for $\rho$ and $p$ as 
\begin{equation} \label{pandrho2}
p(t)=\frac{-\alpha_1}{2\epsilon(4\pi+\mu)t^2}-V(t) ~~~,~~~\rho(t)=\frac{-\alpha_1}{2\epsilon(4\pi+\mu)t^2}+V(t) . 
\end{equation}
In comparison to the first hyperbolic model, Similar behavior has been obtained for different parameters in the hybrid model. For $\epsilon=+1$, $p_{\phi}$ changes sign from positive to negative indicating a cosmic transit. $V(\phi)$, $V(t)$ and $\rho_{\phi}$ are $>0$ where both $V(t)$ and $\rho_{\phi}$ $\rightarrow \infty$ as $t \rightarrow 0$. For $\epsilon=-1$, $p_{\phi}$ is always positive while $\rho_{\phi}$ takes negative values when $\omega_{\phi}<-1$ with a negative scalar potential $V$.
\begin{figure}[H] \label{tap2}
  \centering             
  \subfigure[$q$]{\label{F11}\includegraphics[width=0.29\textwidth]{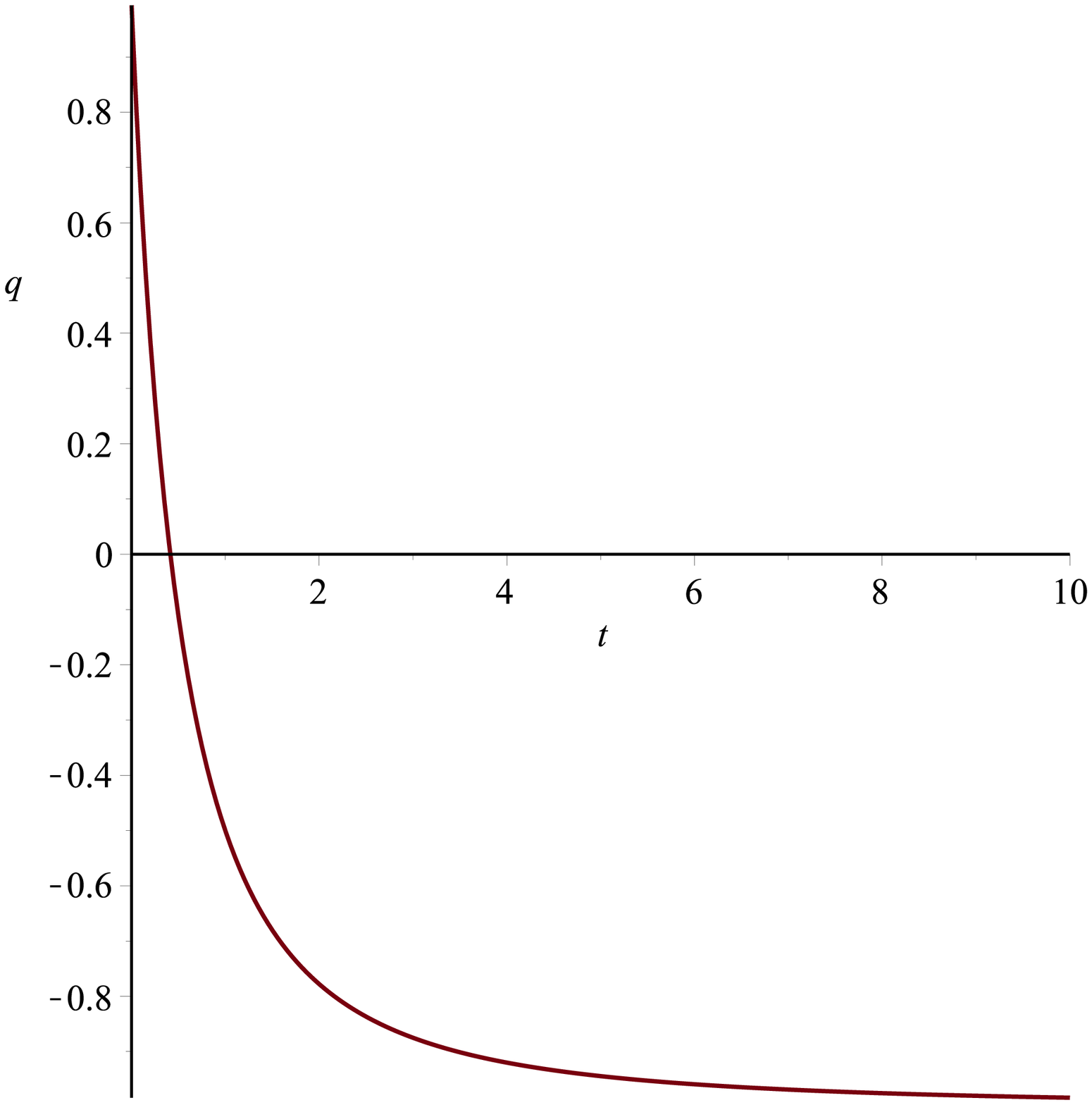}} 
	\subfigure[$j$]{\label{F12}\includegraphics[width=0.29\textwidth]{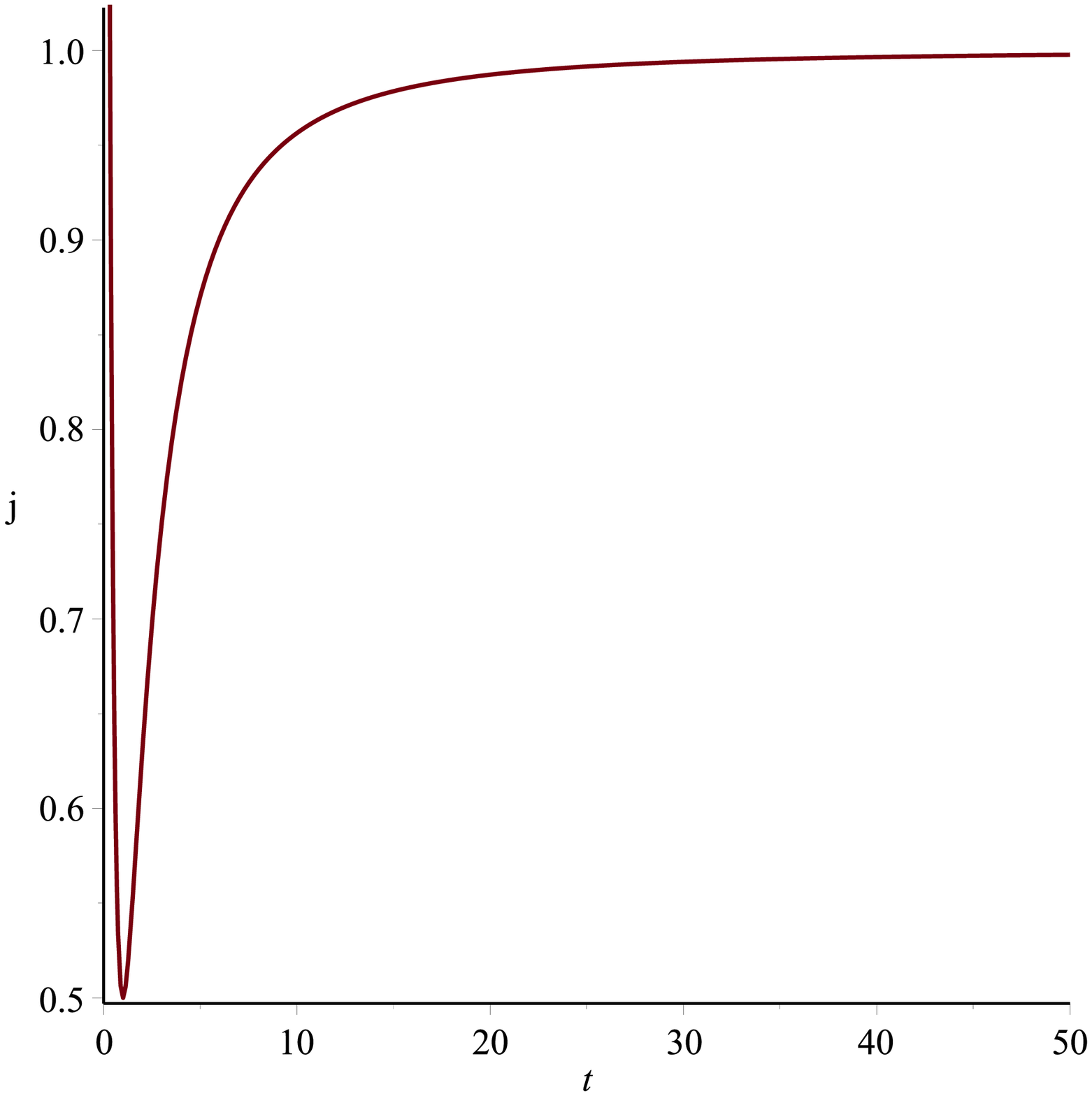}} 
		  \subfigure[$\Lambda(t)$]{\label{F43}\includegraphics[width=0.29\textwidth]{Lambda.eps}} \\
  \subfigure[$p_{\phi}$]{\label{F14}\includegraphics[width=0.29 \textwidth]{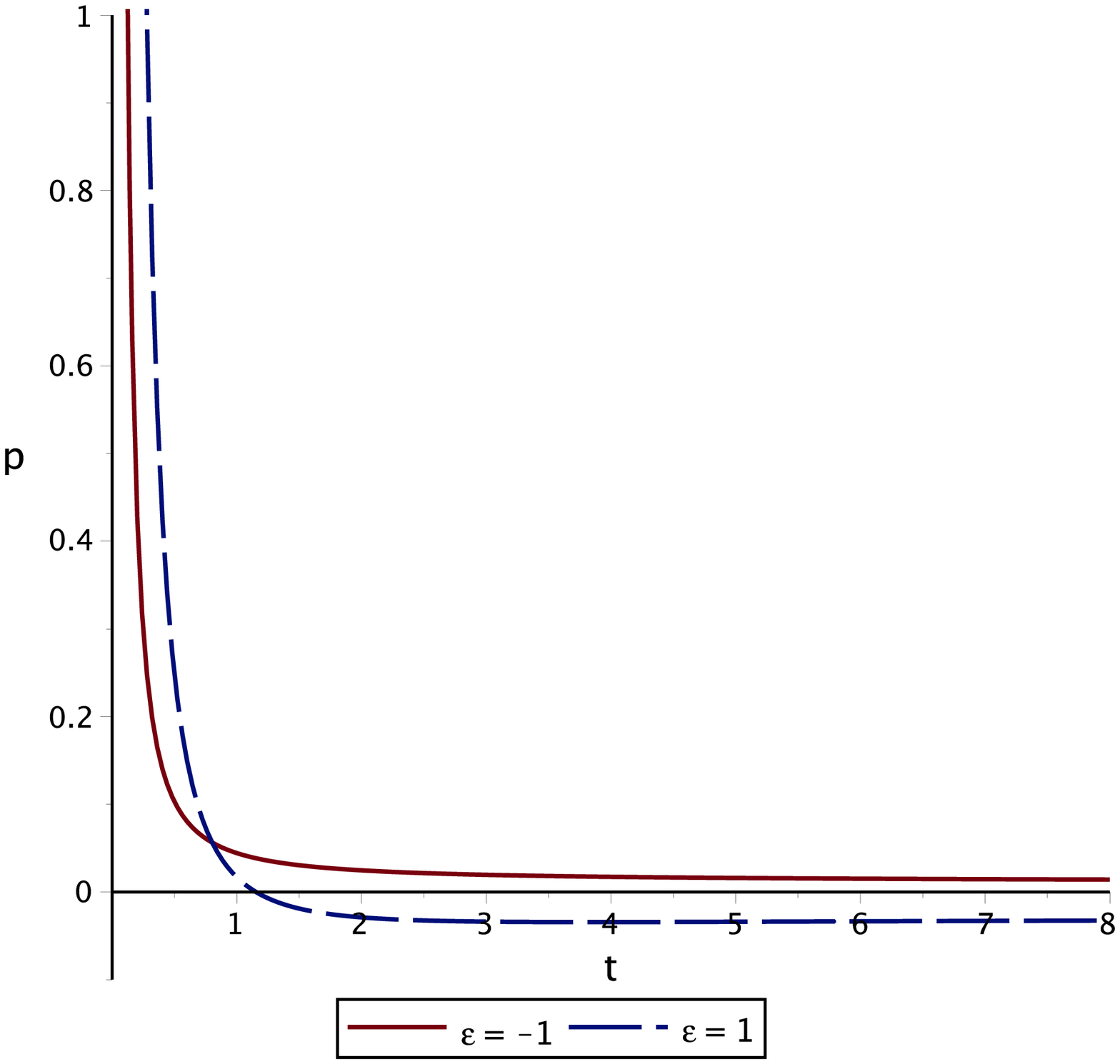}} 
				\subfigure[$\rho_{\phi}$]{\label{F15}\includegraphics[width=0.29 \textwidth]{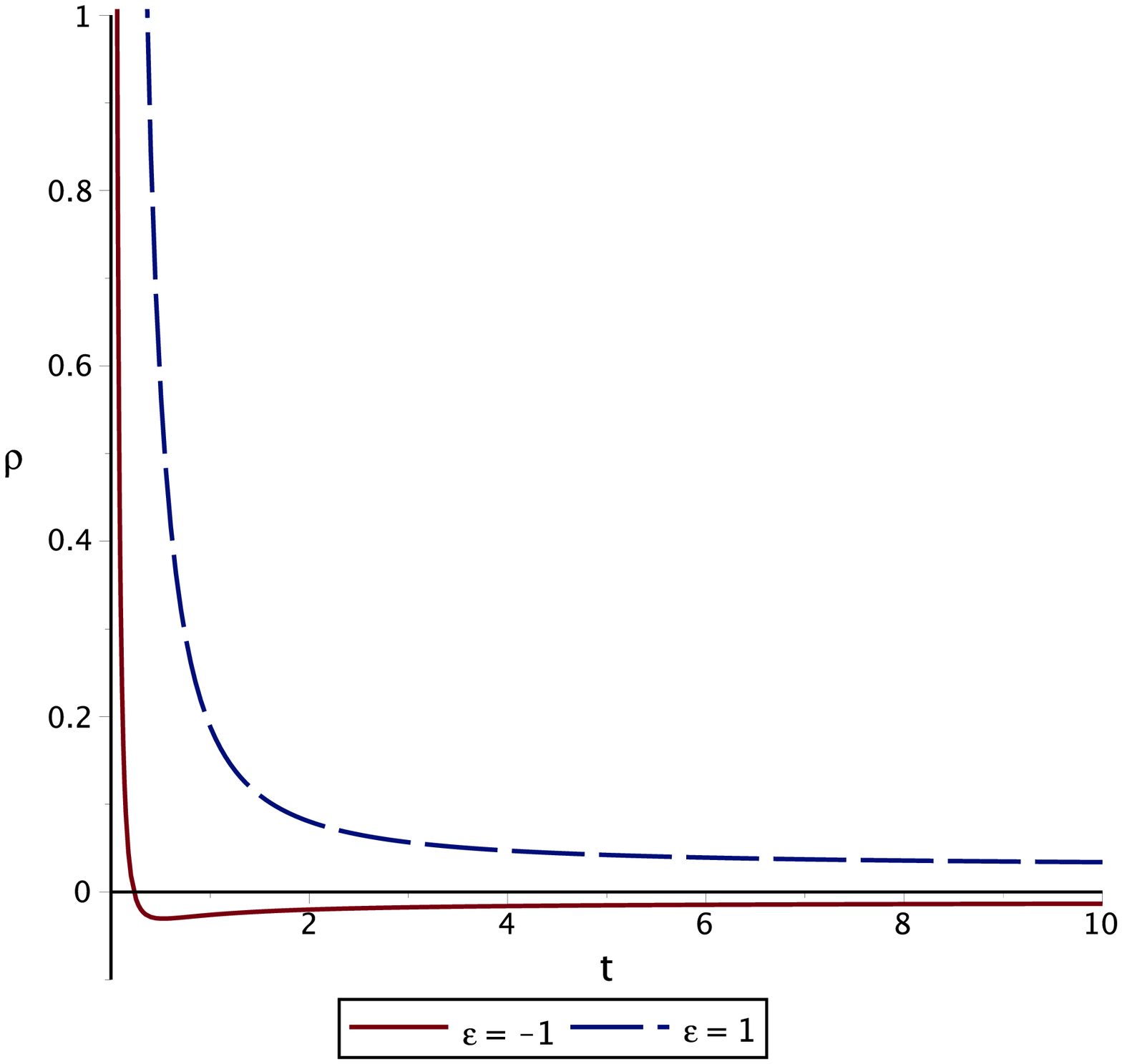}}
	\subfigure[$\omega_{\phi}(t)$]{\label{F170}\includegraphics[width=0.29\textwidth]{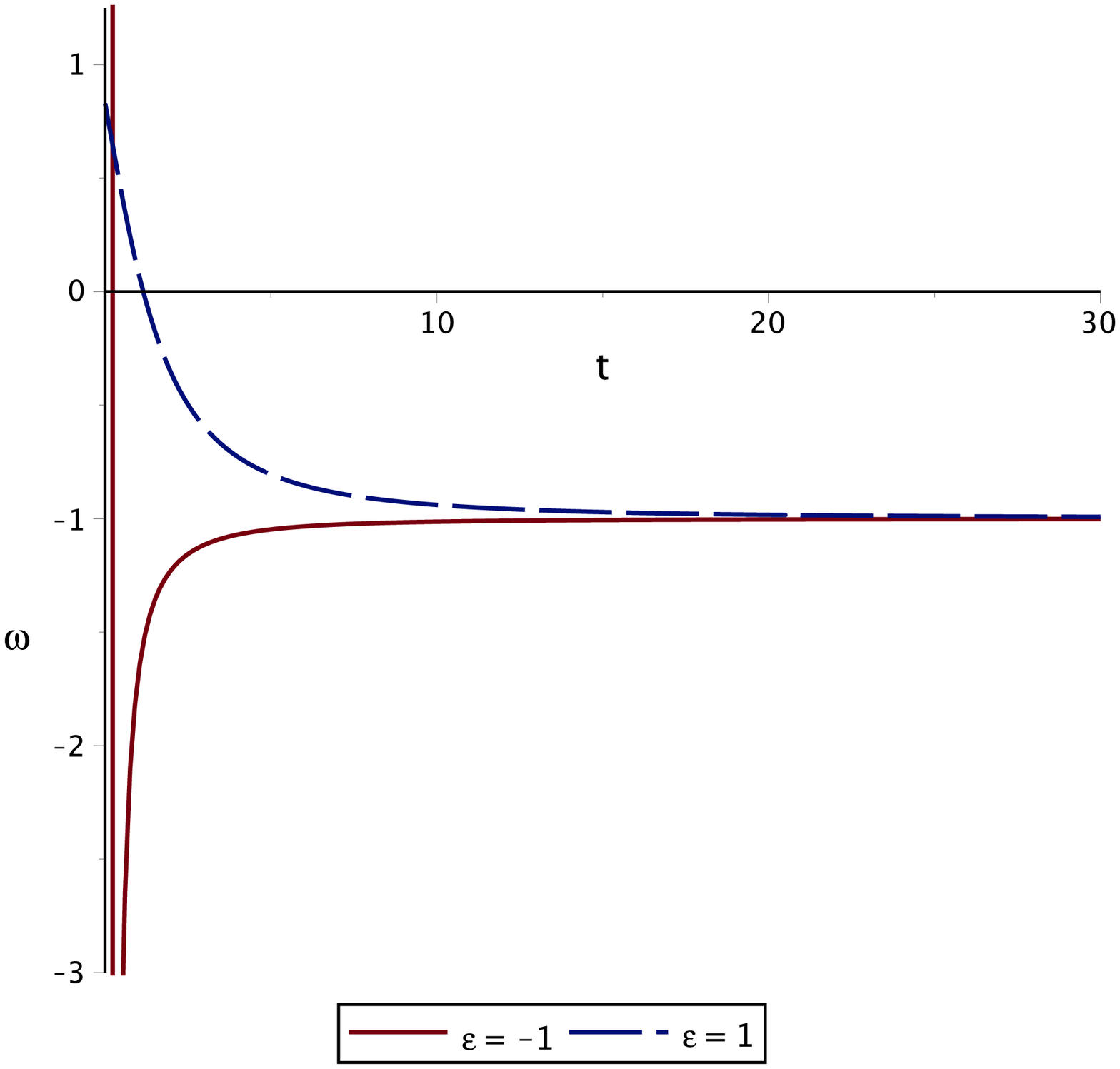}}  \\
	\subfigure[$\phi(t)$]{\label{F17}\includegraphics[width=0.29\textwidth]{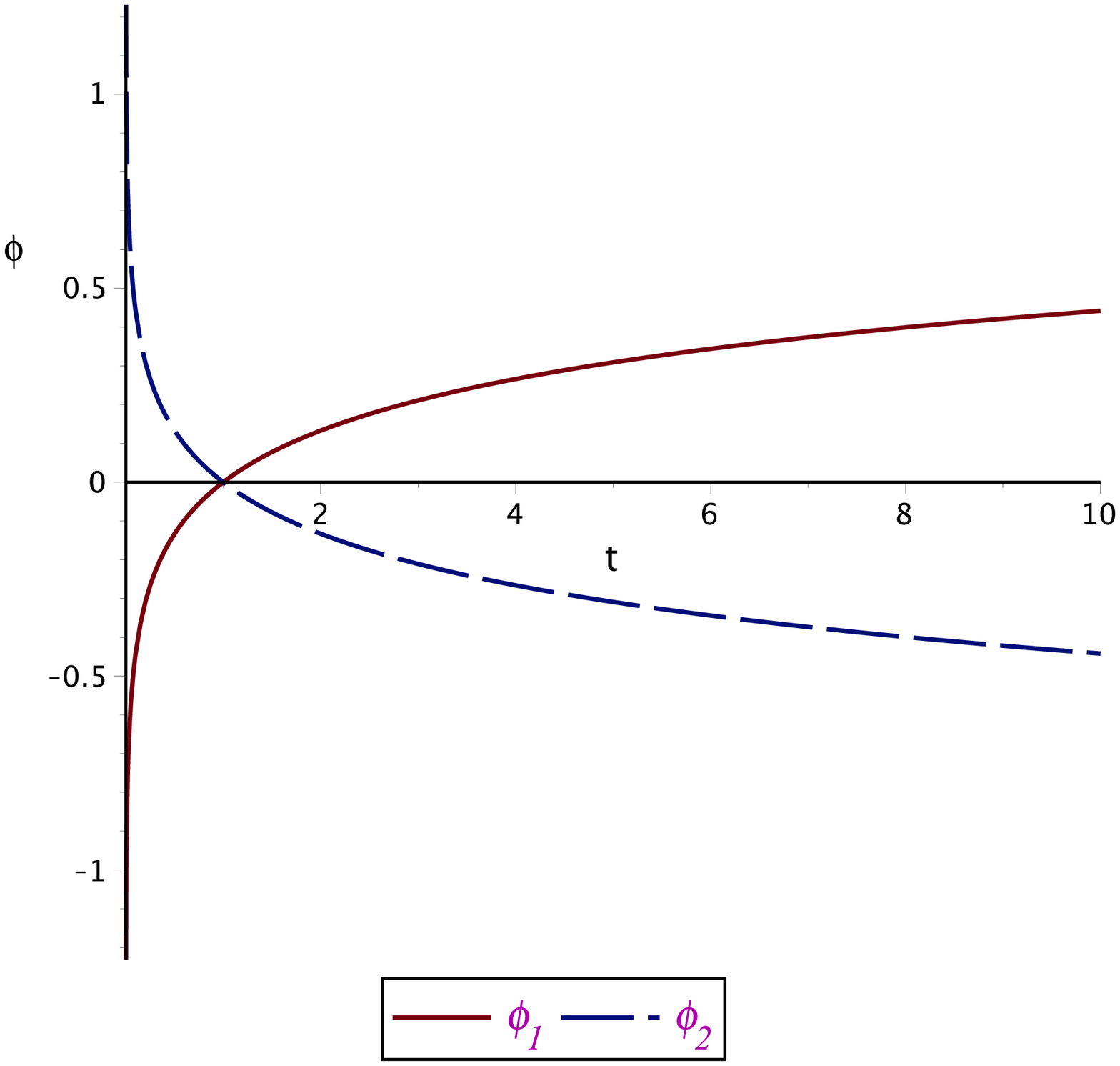}} 
		\subfigure[$V(t)$]{\label{F18}\includegraphics[width=0.29\textwidth]{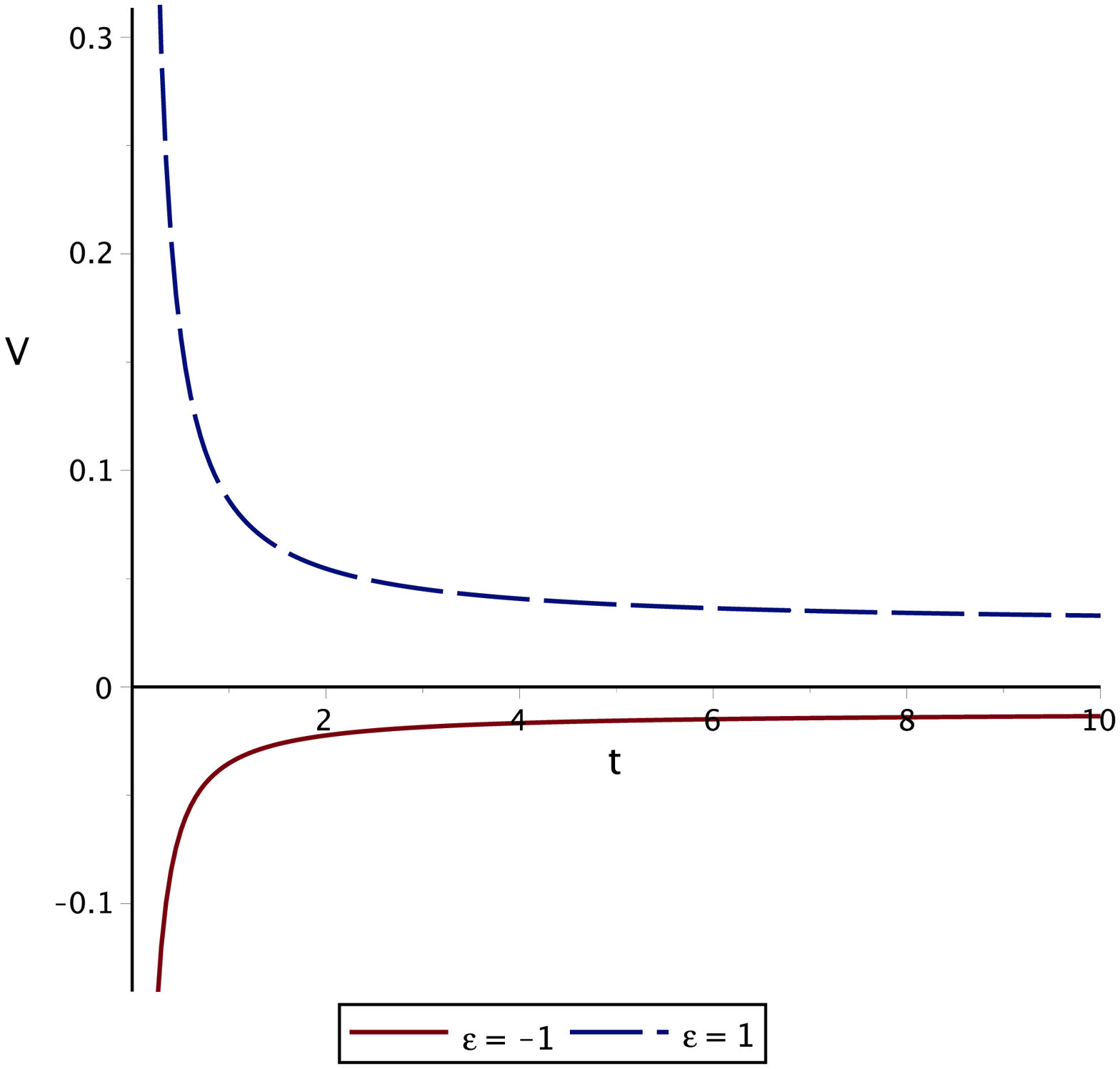}}
			\subfigure[$V(\phi)$]{\label{F19}\includegraphics[width=0.29\textwidth]{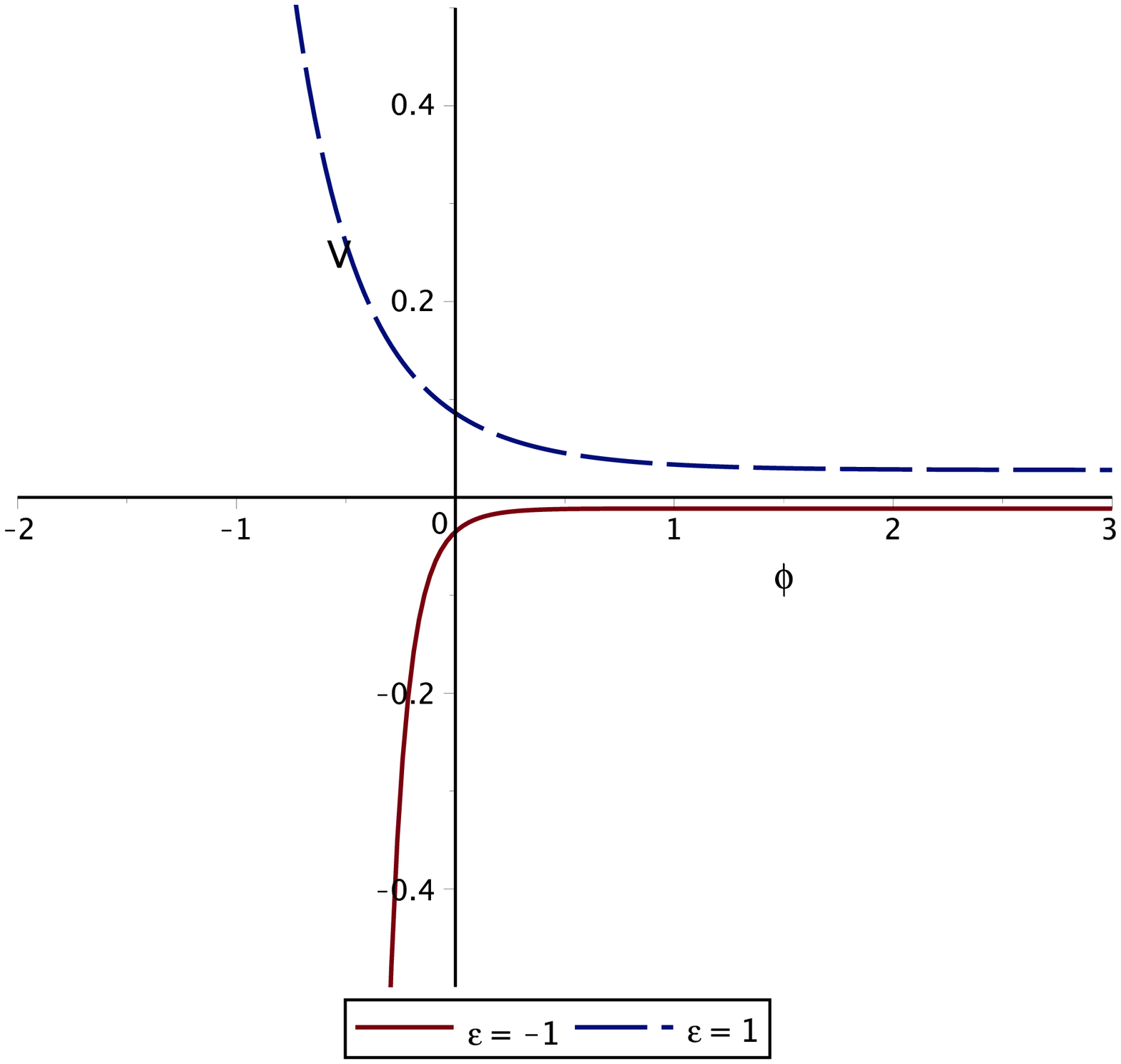}}
   \caption{The second model: (a) A decelerating-accelerating cosmic transit. (b) The jerk parameter $j=1$ at late-times. (c) The cosmological constant reaches a very tiny positive value at the current epoch. (d), (e), \& (f) show $p_{\phi}$, $\rho_{\phi}$ and $\omega_{\phi}$ for $\epsilon=\pm 1$. For the phantom case, the energy density $\rho_{\phi}<0$ when $\omega_{\phi}<-1$. (g) The two solutions of $\phi(t)$ obtained in (\ref{phii}). (h) The scalar potential evolution with time. (g) scalar potential $V$ verses $\phi$ . Here $\alpha_1=\beta_1=0.5$, $\eta=1, \phi_0=0, A=\lambda=\beta=\alpha=0.1$, $\mu=15$ for $\epsilon=-1$ and $-15$ for $\epsilon=1$.}\label{fig:2}
\end{figure}
\begin{figure}[H] \label{tap128}
  \centering             
  \subfigure[$\epsilon=+1$]{\label{p1108}\includegraphics[width=0.29\textwidth]{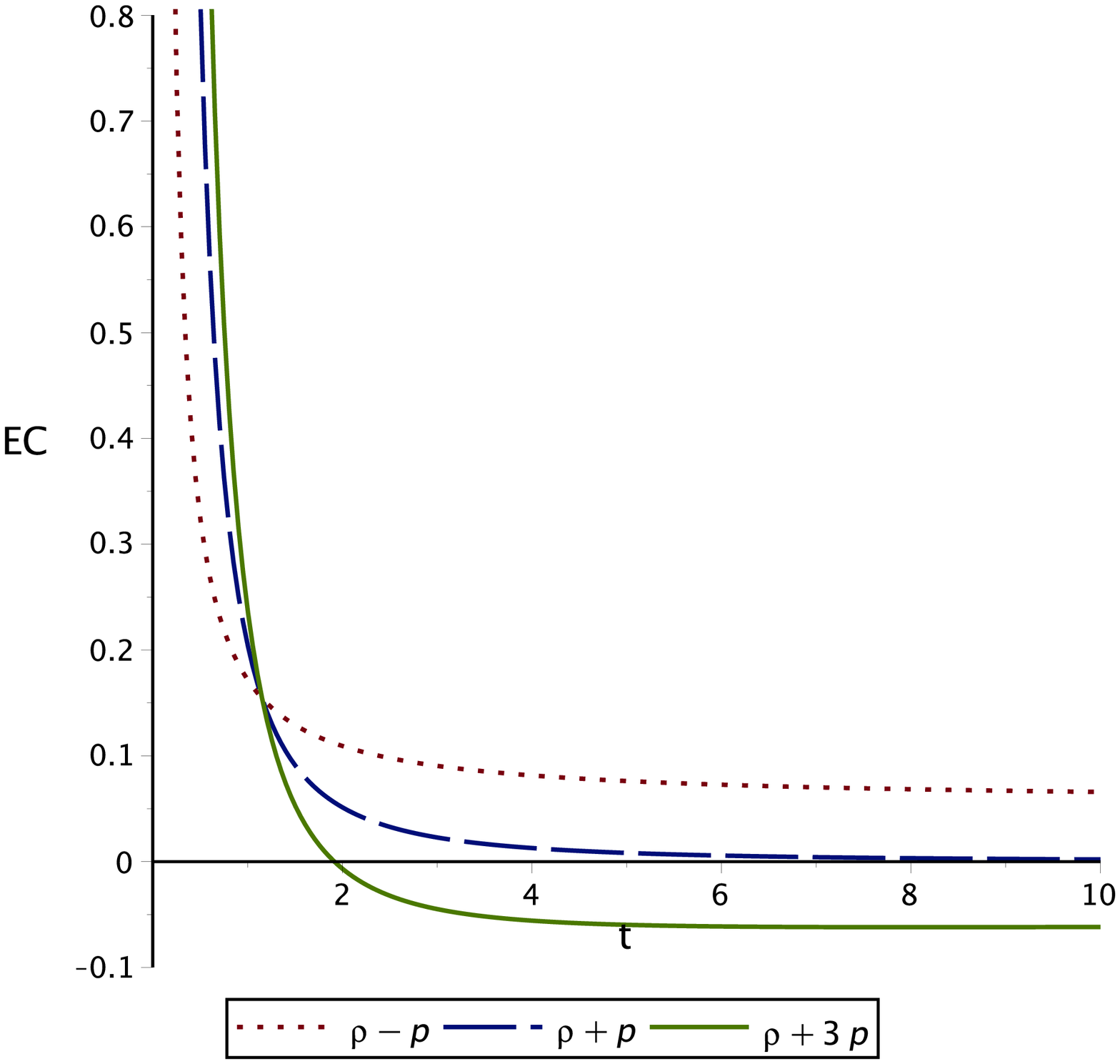}} 
		\subfigure[$\epsilon=-1$]{\label{p1258}\includegraphics[width=0.29\textwidth]{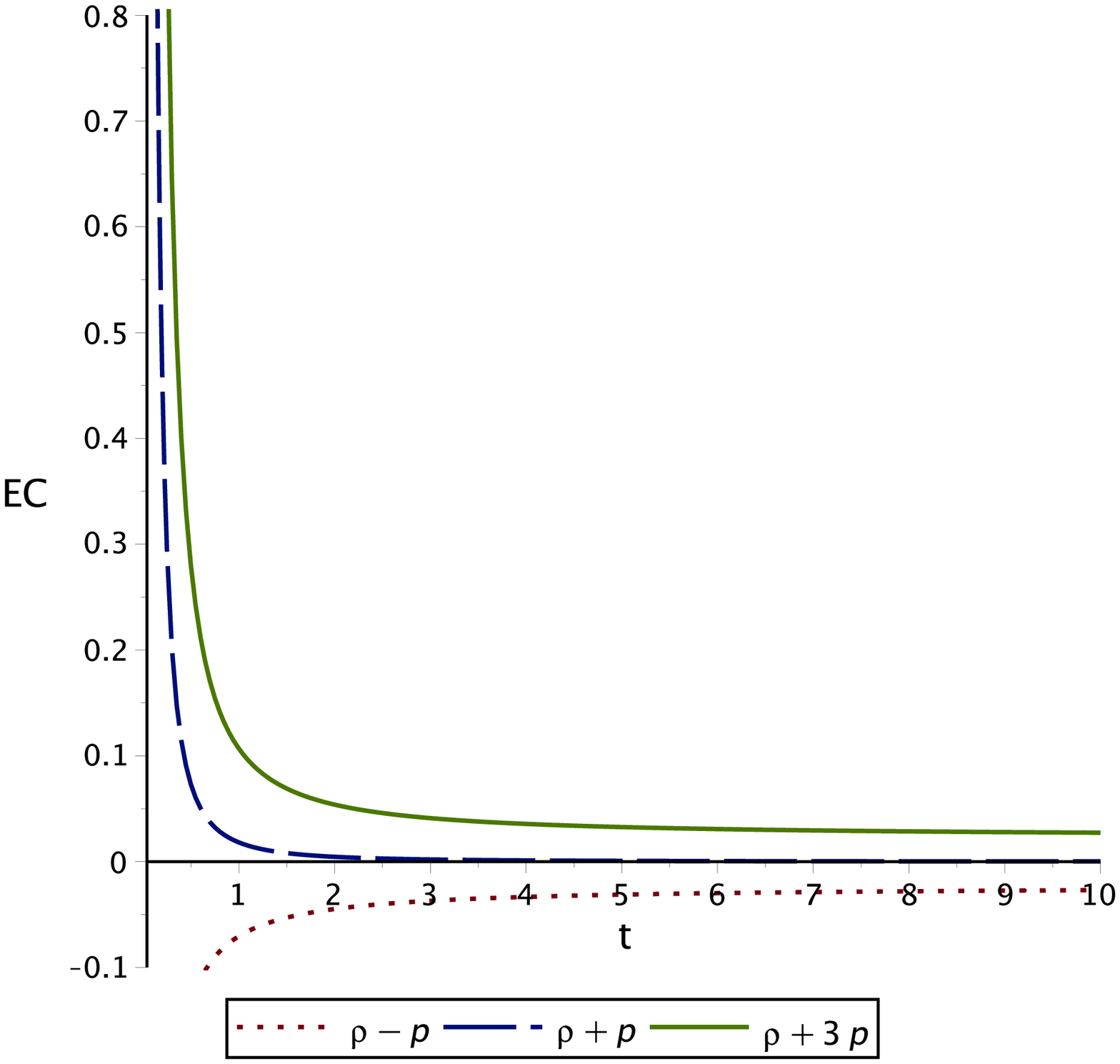}} 
	\subfigure[$dp/dt$]{\label{p1208}\includegraphics[width=0.29\textwidth]{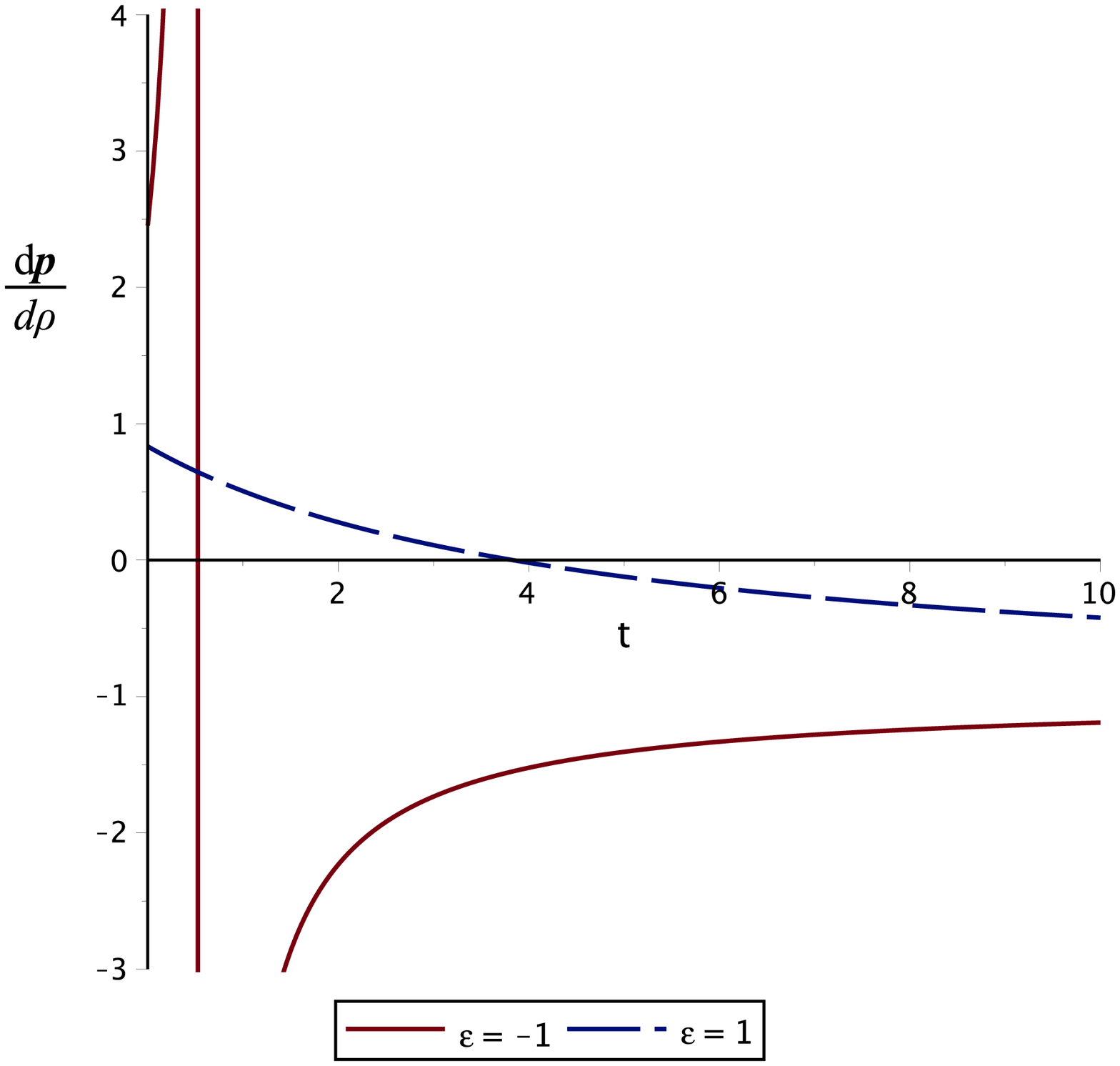}} 
  \caption{ECs and sound speed for the hybrid model. Negative sound speed for the phantom field.}
  \label{fig:13}
\end{figure}
In the current work, we argue that the WEC is not violated for the two models considered with an instability at late-times for the second model which now can be seen in Figure 4(c). The WEC, asserting that the total energy density $\rho$ must be non-negative, is challenged by the notion that a negative term in the energy density can coexist if the overall energy density remains positive. Figure 4(c) shows that the sound speed causality condition is satisfied only within a specific time interval (for late-times) for a normal scalar field while it is always violated for the phantom field. The phantom field, for both the hyperbolic and hybrid models, has a positive pressure $p_{\phi}>0$ and a negative scalar potential $V(\phi)$. Also, Its energy density $\rho_{\phi}=E_k+V$ takes negative values when the equation of state parameter $\omega_{\phi}<-1$. Figure 4(b) shows that  $p_i+\rho_i \geq 0$ for both normal and phantom fields.

\section{Conclusion}
We have revisited  the scalar field cosmology in $f(R,T)$ gravity through two models. The main points can be summarized as follows:
\begin{itemize}
\item The evolution of the deceleration parameter indicates that a decelerating-accelerating cosmic transit exists in both models . The jerk parameter also tends to $1$ at late-times where the model tends to a flat $\Lambda$CDM model.
\item The evolution of the varying cosmological constant in both models shows that it tends to a tiny positive value at the present epoch.  
\item The scalar field pressure $p_{\phi}$ in both models shows a sign flipping from positive to negative for normal scalar field  $\epsilon=+1$ , but it's always positive for the phantom field  $\epsilon=-1$ . 
\item  In both models, the scalar potential $V(\phi)>0$ for $\epsilon=+1$ and $<0$ for $\epsilon=-1$ .
\item For the normal field, $\rho_{\phi}>0$ with no crossing to the phantom divide line for $\omega_{\phi}$. For the phantom field we have $\rho_{\phi}<0$ when $\omega_{\phi}<-1$ . 
\item Classical energy conditions have been tested for both cases. For the hyperbolic model, the sound speed causality condition $0 \leq \frac{dp}{d\rho} \leq 1$  is valid only for $\epsilon=+1$. For the hybrid model, this condition is satisfied only for a specific interval of time for the normal scalar field.   
\end{itemize}

\end{document}